\documentclass[authoryear,3p,times]{elsarticle}
\usepackage{graphicx}
\usepackage{amssymb}
\usepackage{multirow,color,supertabular}
\usepackage{longtable}

\hfuzz=\maxdimen
\begin{document}

\begin{frontmatter}
\title{Correlation structure and principal components in global crude oil market}
%Economic Information in the Correlation Matrix of the Global Crude Oil Markets
\author[BS]{Yue-Hua Dai}
\author[BS,SS]{Wen-Jie Xie}
\author[BS]{Zhi-Qiang Jiang}
\author[WSU]{George J. Jiang}
\ead{george.jiang@wsu.edu} %
\author[BS,SS]{Wei-Xing Zhou\corref{cor1}}
\cortext[cor1]{Corresponding author. Address: 130 Meilong Road, P.O. Box 114, School of Business,
              East China University of Science and Technology, Shanghai 200237, China, Phone: +86 21 64253634.}
\ead{wxzhou@ecust.edu.cn} %
%\ead[url]{http://rce.ecust.edu.cn/}%

\address[BS]{School of Business, East China University of Science and Technology, Shanghai 200237, China}
%\address[RCE]{Research Center for Econophysics, East China University of Science and Technology, Shanghai 200237, China}
\address[SS]{Department of Mathematics, East China University of Science and Technology, Shanghai 200237, China}
\address[WSU]{Department of Finance and Management Science, Washington State University, Pullman, U.S.A.}
%\address[ICCT]{Key Laboratory of Coal Gasification and Energy Chemical Engineering (MOE), East China University of Science and Technology, Shanghai 200237, China}

\begin{abstract}
  This article investigates the correlation structure of the global crude oil market using the daily returns of 71 oil price time series across the world from 1992 to 2012. We identify from the correlation matrix six clusters of time series exhibiting evident geographical traits, which supports \cite{Weiner-1991-TEJ}'s regionalization hypothesis of the global oil market. We find that intra-cluster pairs of time series are highly correlated while inter-cluster pairs have relatively low correlations. Principal component analysis shows that most eigenvalues of the correlation matrix locate outside the prediction of the random matrix theory and these deviating eigenvalues and their corresponding eigenvectors contain rich economic information. Specifically, the largest eigenvalue reflects a collective effect of the global market, other four largest eigenvalues possess a partitioning function to distinguish the six clusters, and the smallest eigenvalues highlight the pairs of time series with the largest correlation coefficients. We construct an index of the global oil market based on the eigenfortfolio of the largest eigenvalue, which evolves similarly as the average price time series and has better performance than the benchmark $1/N$ portfolio under the buy-and-hold strategy.

\end{abstract}


\end{frontmatter}

%main text
\section{Introduction}
\label{S1:Introduction}

Crude oil is the life blood of our modern industrial society and a unique strategic resource that is of crucial importance to any economy. The prices of crude oil are driven by the supply/demand imbalance and the uncertainty of this imbalance which causes increased speculations \citep{AlvarezRamirez-Cisneros-IbarraValdez-Soriano-2002-PA,He-Fan-Wei-2009-EE,Kaufmann-Ullman-2009-EE,Sornette-Woordard-Zhou-2009-PA}. There is huge literature devoting to the study of the dynamics of crude oil prices and their mutual relationships.
A considerable portion of the literature focuses on the co-movement and convergence of oil prices at different regions. \cite{Adelman-1984-TEJ} asserts that the global market of crude oils is unified as ``one great pool''. Conversely, \cite{Weiner-1991-TEJ} argues that the crude oil markets are regionalized which challenges the ``one great pool'' hypothesis of \cite{Adelman-1984-TEJ}. These two competing hypotheses have stimulated many debates and extensive studies \citep{Rodriguez-Williams-1993-ESR,Weiner-1993-ESR,Rodriguez-Williams-1994-ESR}.

The majority of empirical studies support the ``one great pool'' hypothesis. In this line, different econometric methods have been adopted and different ``definitions'' of market unification have been implicitly assumed. \cite{Sauer-1994-TEJ} incorporates cointegration relationships into multivariate time series models to examine the extent of regionalization in the global market for crude oil imports and finds that the empirical results support \cite{Adelman-1984-TEJ}'s ``one great pool'' assertion. \cite{Gulen-1997-TEJ,Gulen-1999-TEJ} performs cointegration tests for co-movement of monthly and weekly prices on three groups of crude oil of similar quality and finds that the world crude oil market is unified over the 1980-95 period, rejecting the regionalization hypothesis. \cite{Bentzen-2007-AE} finds bidirectional causality among these major crude oil prices (OPEC, Brent and WTI) and argues that the regionalization hypothesis of the global oil market is thus rejected. \cite{Fattouh-2010-EE} investigates the dynamics of crude oil price differentials using the two-regime threshold autoregressive (TAR) method of \cite{Caner-Hansen-2001-Em} and finds strong evidence of threshold effects in the adjustment process to the long-run equilibrium. Since the crude oil prices are linked, \cite{Fattouh-2010-EE} argues that the oil market is ``one great pool'' at the very general level. \cite{Reboredo-2011-EM} examines the dependence structure between crude oil benchmark prices using copulas and finds evidence of significant symmetric upper and lower tail dependence between crude oil prices. He states that crude oil prices are linked with the same intensity during bull and bear markets, thus favoring the ``one great pool'' hypothesis over the regionalization hypothesis.

\cite{Kaufmann-Ullman-2009-EE} argue there is no room for innovations in world oil prices to enter the market if the world oil market is unified and there would be no causal relationships between prices for different crude oils. They point out that changes may first appear in the price of one or more benchmark crude oils and subsequently spread through the global market. They find evidence of Granger causality from benchmark markers to other crude oil markets. \cite{Akhmedjonov-Lau-2012-EM} study the monthly energy prices of four energy products for 83 Russian regions using the Exponential Smooth Auto-Regressive Augmented Dickey-Fuller unit root test and find no evidence of a fully integrated national energy market in Russia. \cite{Liu-Chen-Wan-2013-EM} examine the regionalization issue by investigating the integration between China's and four major crude oil markets with a threshold error correction model and find only unidirectional volatility spillover running from benchmark markets to China's oil market. Their results do not favor the ``one great pool'' hypothesis.

Our work contributes to this literature by uncovering the correlation structure of the global crude oil market with principal component analysis \citep{Jolliffe-2002} and random matrix theory \citep{Mehta-2006}. The principal component analysis has been widely applied in finance \cite[see,~for~example,][and~references~therein]{Kritzman-Li-Page-Rigobon-2011-JPM,Billio-Getmansky-Lo-Pelizzon-2012-JFE}. However, principal component analysis is less adopted in the studies of energy markets. \cite{Chantziara-Skiadopoulos-2008-EE} perform principal component analysis on the prices of crude oil, heating oil and gasoline on the New York Mercantile Exchange and crude oil futures on the International Petroleum Exchange and show that retained principal components have limited power in predicting the prices.

On the other hand, random matrix theory has been applied extensively in studying multiple financial time series \citep{Laloux-Cizean-Bouchaud-Potters-1999-PRL,Plerou-Gopikrishnan-Rosenow-Amaral-Stanley-1999-PRL,Kwapien-Drozdz-2012-PR}. Random matrix theory is in essence equivalent to principal component analysis because both of them deal with the correlation matrix and its eigenvalues. Under the framework of random matrix theory, if the eigenvalues of the real time series differ from the prediction of random matrix theory, there must exist hidden economic information in those deviating eigenvalues. For stock markets, there are several deviating eigenvalues in which the largest eigenvalue reflects a collective effect of the whole market and other largest eigenvalues can be used to identify industrial sectors \citep{Plerou-Gopikrishnan-Rosenow-Amaral-Guhr-Stanley-2002-PRE} or clusters of stocks with strong cross-correlations \citep{Shen-Zheng-2009a-EPL}. Different information can be extracted for housing markets \cite{Meng-Xie-Jiang-Podobnik-Zhou-Stanley-2014-SR} and global stock markets \cite{Song-Tumminello-Zhou-Mantegna-2011-PRE}. Such kind of analysis enables us to uncover market driving forces \citep{Shapira-Kenett-Jacob-2009-EPJB} as well as mutual and common influence \citep{Garas-Argyrakis-2007-PA}.

We apply principal component analysis and random matrix theory to investigate the correlation structure of the global crude oil market by using 71 spot price time series from different countries. We are able to identify six clusters of oil price time series that have clear geographic traits. This finding supports the regionalization hypothesis of \cite{Weiner-1991-TEJ}. We also extract rich economic information from the deviating eigenvalues, suggesting that the global crude oil market has a very distinct correlation structure. This rest of this paper is organized as follows. Section \ref{S1:Data} describes the data sets and presents the summary statistics. Section \ref{S1:Correlation} investigates the cross-correlation structure of the global crude oil market. Section \ref{S1:Eigen} studies the correlation matrix and explores the economic information contents embedded in the principal components and the smallest eigenvalues. We conclude in Sec.~\ref{S1:Conclusion}.

\section{Data description}
\label{S1:Data}

\subsection{Data set}

We retrieved from the Bloomberg database 71 daily spot price time series of crude oil in various markets all over the world. The spot price time series cover a period from October 1992 to December 2012. These crude oil price time series differ in several aspects. The oil markets may locate in different countries or regions including the main crude oil export countries such as Iran and Saudi Arab or different places in a country. Particularly, some oil price series are recorded according to their different crude oil qualities (including density and sulfur content). For the original data, we intersect them and then complement price of the missing data same as that of the previous day. If data of the previous day is still missing, we complement them with that of the day before previous day, the rest is done in the same manner. The labels of different crude oils, their corresponding ticker names, and the length of time series are presented in Table \ref{TB:Oil:Datasets}.

\subsection{Returns}

The logarithmic return of the $i$th crude oil price series over a time scale $\Delta t$ is calculated as follows:
\begin{equation}
 r_{i}(t)=\ln P_{i}(t)-\ln P_{i}(t-\Delta{t}),
 \label{Eq:Return:Defintion}
\end{equation}
where $P_{i}(t)$ denotes the price of $i$th crude oil at time $t$. Note that the label $i$ for each time series is given in the first column of Table~\ref{TB:Oil:Datasets}. In this work, we present the results for daily returns with $\Delta{t}=1$ day. To take into account the nonsynchronous trading of crude oils all over the world, we repeated the analysis for the weekly returns with $\Delta{t}=7$ days. The results for weekly returns are qualitatively the same as the daily data. In our analysis, abnormally large price fluctuations ($>40\%$) are removed. However, the inclusion of these large fluctuations does not change the main results.

Table \ref{TB:Oil:Datasets} presents the summary statistics of the daily returns for each time series. Usually, if the maximum return (Max) is large if the absolute minimum return (Min) is large for a time series. All the mean returns are positive and have an order of 0.02\% to 0.05\%. The standard deviation of returns is about 0.023. The return distributions of most time series are right-skewed or left-skewed. We find that the excess kurtosis of each return time series is significantly great than 3, the kurtosis of normal distributions, indicating that the return distributions have fat tails.

\begin{figure}
  \centering
  \includegraphics[width=8.3 cm]{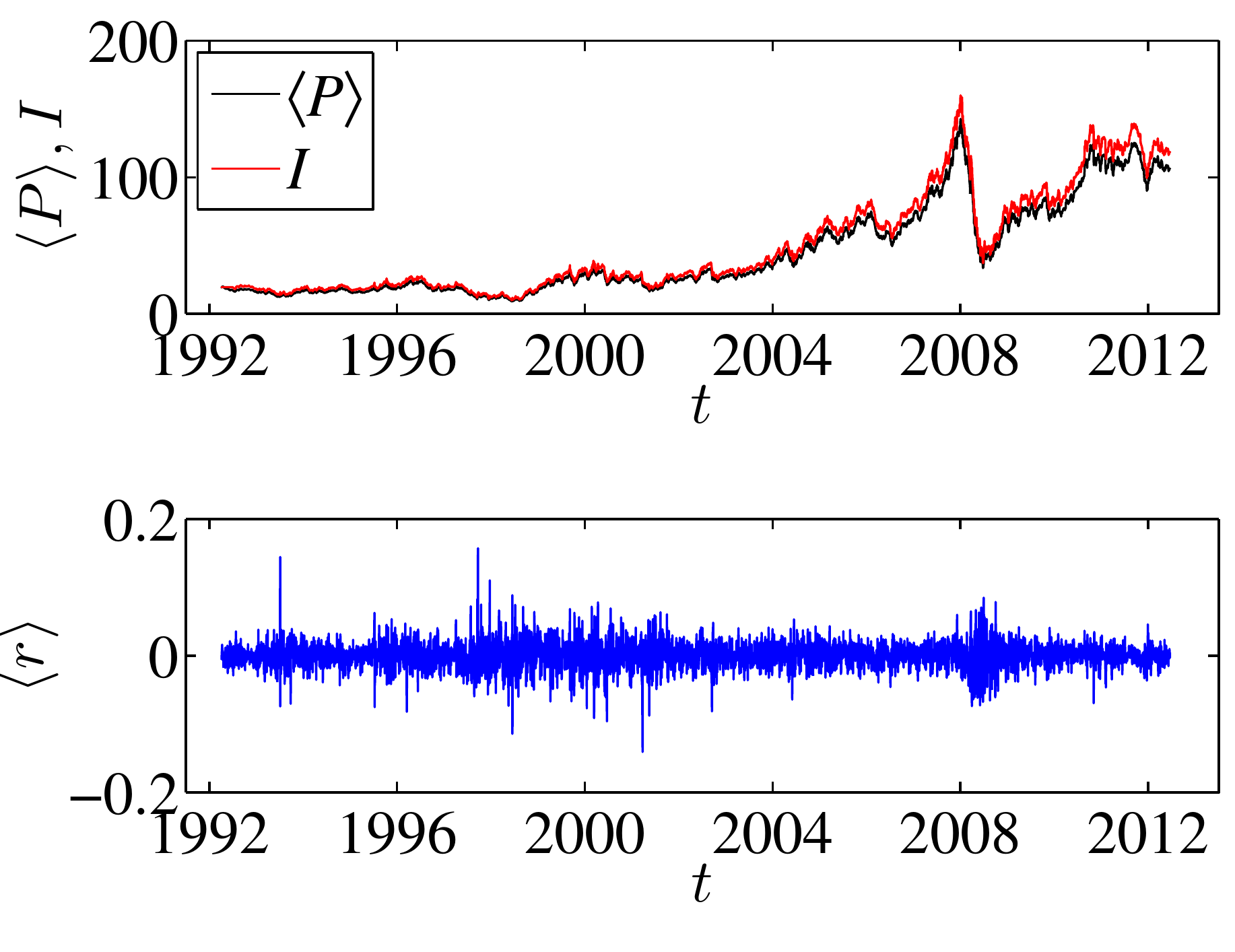}
  \caption{(Color online) Evolution of average daily prices, the constructed index based on the eigenportfolio of the largest eigenvalue, and average logarithmic returns of crude oils.}\label{Fig:PCA:Oil:Mean:Price:Return}
\end{figure}

In Fig.~\ref{Fig:PCA:Oil:Mean:Price:Return}, we show the evolution of the average oil price $\langle{P(t)}\rangle=\frac{1}{71}\sum_{i=1}^{71}P_i(t)$ and the corresponding average logarithmic return $\langle{r(t)}\rangle=\frac{1}{71}\sum_{i=1}^{71}r_i(t)$. The most significant pattern in the price evolution is the boom and bust of a huge bubble around 2008, mainly caused by speculations due to market uncertainty \citep{Sornette-Woordard-Zhou-2009-PA}. The time series of the average returns evidently exhibits the volatility clustering phenomenon. We also find that the distribution of the average returns is leptokurtic with the excess kurtosis being 10.02 and left skewed with the skewness being -0.091.

\section{Cross correlation structure}
\label{S1:Correlation}

\subsection{Correlation matrix}

We calculate the pairwise cross-correlation coefficients between any two crude oil return time series. For simplicity, the original returns for each crude oil time series are standardized as follows:
\begin{equation}
  g_{i}(t)=\frac{r_i(t)-\langle r_i(t)\rangle}{\sigma_i},
  \label{Eq:NormalizedReturn:Defintion}
\end{equation}
where $\langle\cdot\rangle$ denotes the time average of a given time series and $\sigma_i=\sqrt{\langle r_i(t)^2\rangle-\langle r_i(t)\rangle^2}$ is the standard deviation of $r_i(t)$. The cross-correlation coefficients $c_{ij}$ are computed as follows:
\begin{equation}
 c_{ij}=\langle g_i(t)g_j(t)\rangle.
 \label{Eq:Corrcoef:Defintion}
\end{equation}
By definition, $c_{ij}$ ranges from -1 to 1, where $c_{ij}=1$ corresponds to perfect positive cross-correlations, $c_{ij}=-1$ reflects perfect negative cross-correlations, and $c_{ij}=0$ indicates no cross-correlations between $r_i(t)$ and $r_j(t)$.

\begin{figure*}
  \centering
  \includegraphics[width=8cm,height=6cm]{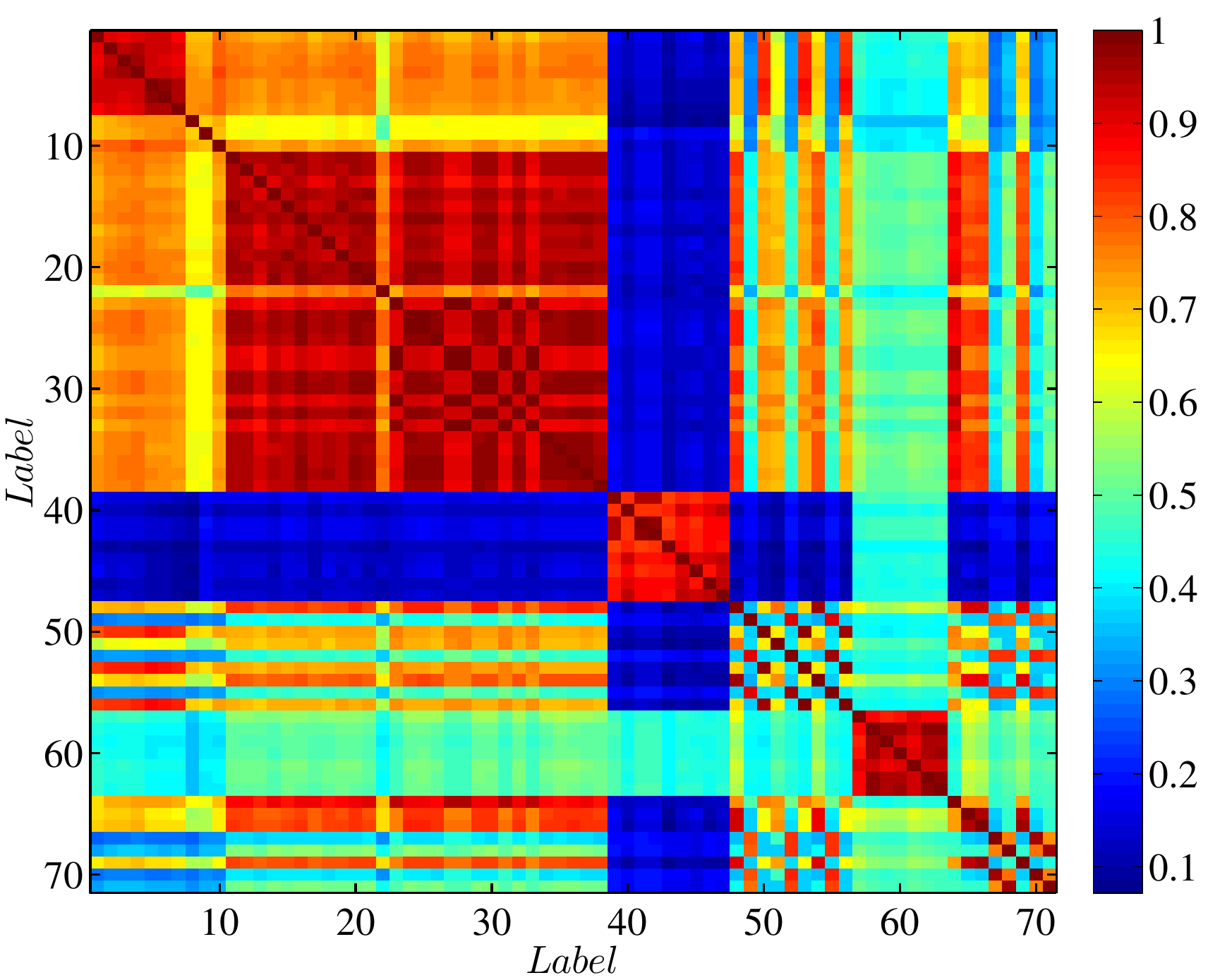}
  \includegraphics[width=8cm,height=6cm]{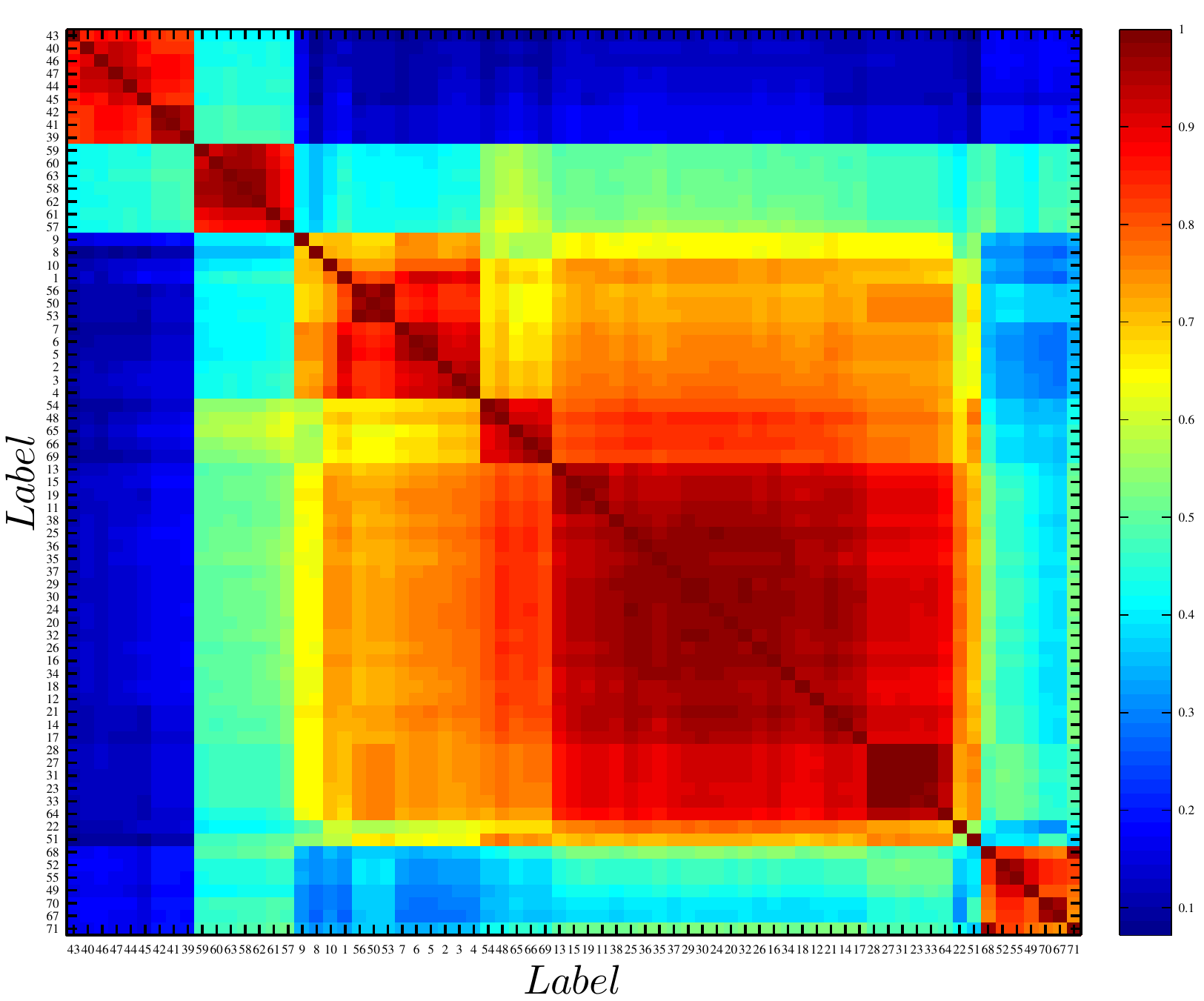}
  \includegraphics[width=16cm]{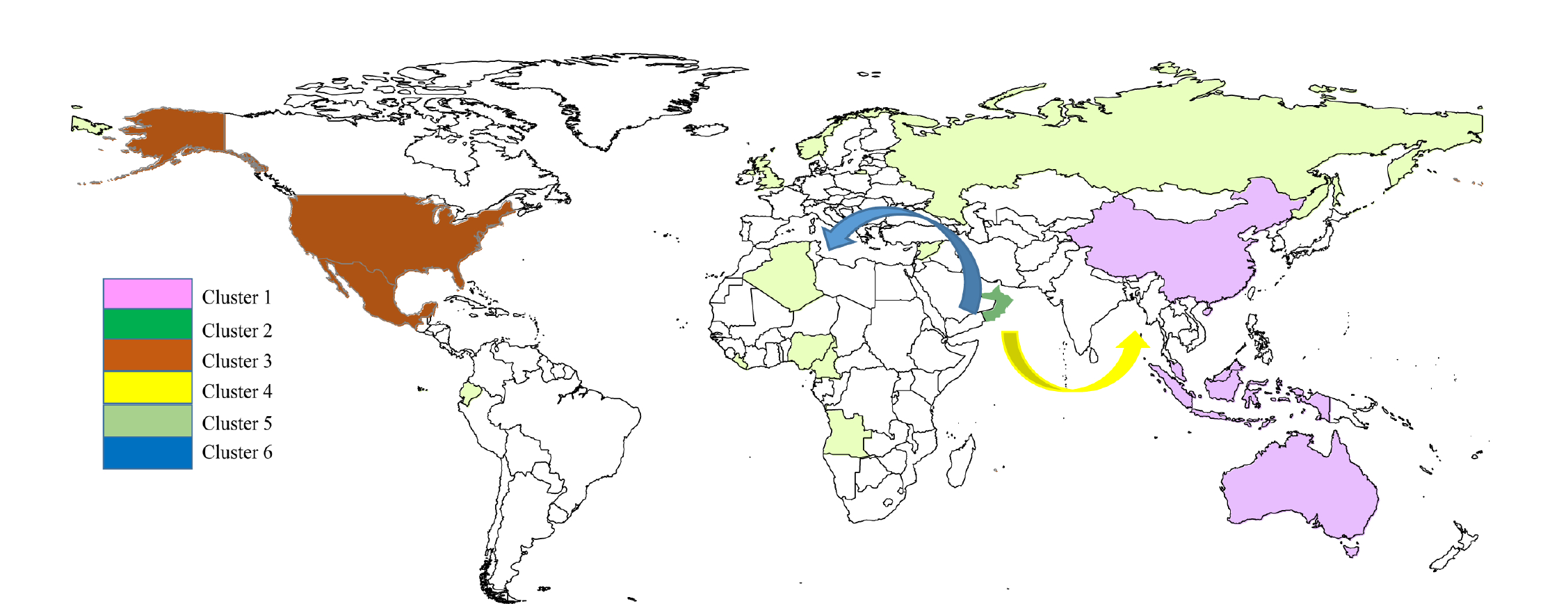}
  \vskip  -12.2cm   \hskip   -15.5cm {\textsf{(a)}}
  \vskip  -0.45cm   \hskip    0.10cm {\textsf{(b)}}
  \vskip   5.9cm   \hskip   -15.45cm {\textsf{(c)}}
  \vskip  5.5cm
  \caption{(Color online) Cross-correlation structure among 71 crude oil price time series all over the world. (a) Correlation matrix of the daily logarithmic returns. (2) Clusters identification of the oil price time series using the box clustering algorithm and the consensus clustering method. (c) Colored map for the six clusters.}
  \label{Fig:PCA:Oil:Cij}
\end{figure*}

Fig.~\ref{Fig:PCA:Oil:Cij}(a) shows the resulting correlation matrix. The correlation structure of the global crude oil market exhibits intriguing features. There are dense blocks of very high cross-correlations with $c_{ij}$ close to 1. This feature stems from the geographic closeness of the time series, such as the block around 60 for the Middle East. We also find that crude oil prices in Asia and Pacific region correlate to those in the Middle East, but have very low correlations with the American and European markets. The main reason is that crude oil in Asian and Pacific regions is imported from the Middle East areas with prices being determined by their officials.

During the sample period, the average correlation coefficient $\langle c\rangle =0.57 $ is much higher than most stock markets \citep{Plerou-Gopikrishnan-Rosenow-Amaral-Guhr-Stanley-2002-PRE,Pan-Sinha-2007-PRE,Shen-Zheng-2009a-EPL}, which indicates that the crude oil markets are more vulnerable to risks than stock markets.

\subsection{Identification of clusters}

In order to extract the clusters of time series in an objective way, we adopt modern algorithms for community detection in complex networks. For the correlation matrix $\mathbf{C}$, following the idea of \cite{Lancichinetti-Fortunato-2012-SR} and \cite{Meng-Xie-Jiang-Podobnik-Zhou-Stanley-2014-SR}, we combine the box clustering algorithm of \cite{SalesPardo-Guimera-Moreira-Amaral-2007-PNAS} and the consensus clustering method of \cite{Lancichinetti-Fortunato-2012-SR} to search for clusters of crude oil time series.

We first determine the optimal ordering of $\mathbf{C}$ by identifying the largest elements in $\mathbf{C}$ close to the backward diagonal $i=-j$, where the simulated annealing approach is adopted to minimize the cost function
\begin{equation}
  Q = \sum_{i,j=1}^{71} C_{ij}|i-j|.
\end{equation}
We then use a greedy algorithm to partition clusters of time series \citep{SalesPardo-Guimera-Moreira-Amaral-2007-PNAS}. We repeat this procedure 200 times and obtain 200 partitions. We construct an affinity matrix ${\mathbf{A}}'$ whose element $A_{ij}'$ is the number of partitions in which $i$ and $j$ are assigned to the same cluster, divided by the number of partitions 200. Furthermore, we apply the clustering method to the affinity matrix ${\mathbf{A}}'$, resulting in a final partition ${\mathbf{A}}$ \citep{Lancichinetti-Fortunato-2012-SR}. We finally rearrange the order of states in ${\mathbf{C}}$ to be the same as in ${\mathbf{A}}$ \citep{Meng-Xie-Jiang-Podobnik-Zhou-Stanley-2014-SR}.

The resultant rearranged correlation matrix is illustrated in Fig.~\ref{Fig:PCA:Oil:Cij}(b), in which six clusters of time series are identified. In the last column of Table \ref{TB:Oil:Datasets}, we provide the cluster information of each time series with the clusters labeled 1 to 6 from left to right in Fig.~\ref{Fig:PCA:Oil:Cij}(b). In addition, we color the world map based on the clusters, as shown in in Fig.~\ref{Fig:PCA:Oil:Cij}(c). We assign each cluster a unique color and a region or a country is colored if its crude oil market belongs to a specific cluster.

There are two arrows in the colored map, one in blue and the other in yellow. The blue arrow stands for Cluster 6 where the crude oils exported from Mideast area to Northwest Europe, while the yellow arrow represents Cluster 4 where the crude oils exported from Mideast area to Asia-Pacific area. The rest time series form Cluster 2 in Middle East. Cluster 1 contains the time series in Asia and Australia, which are uncorrelated with all other time series as shown in Fig.~\ref{Fig:PCA:Oil:Cij}(b). Cluster 3 mainly contains time series in North America and Cluster 5 in Europe and Nigeria. As shown in Table \ref{TB:Oil:Datasets}, there are only a few exceptions. Therefore, the correlation structure of the global oil market exhibits remarkable geographical traits, which is reminiscent of the global stock markets \citep{Song-Tumminello-Zhou-Mantegna-2011-PRE} and the US housing market \citep{Meng-Xie-Jiang-Podobnik-Zhou-Stanley-2014-SR}.

\subsection{Distribution of correlation coefficients}

\begin{figure}
  \centering
\includegraphics[width=8cm]{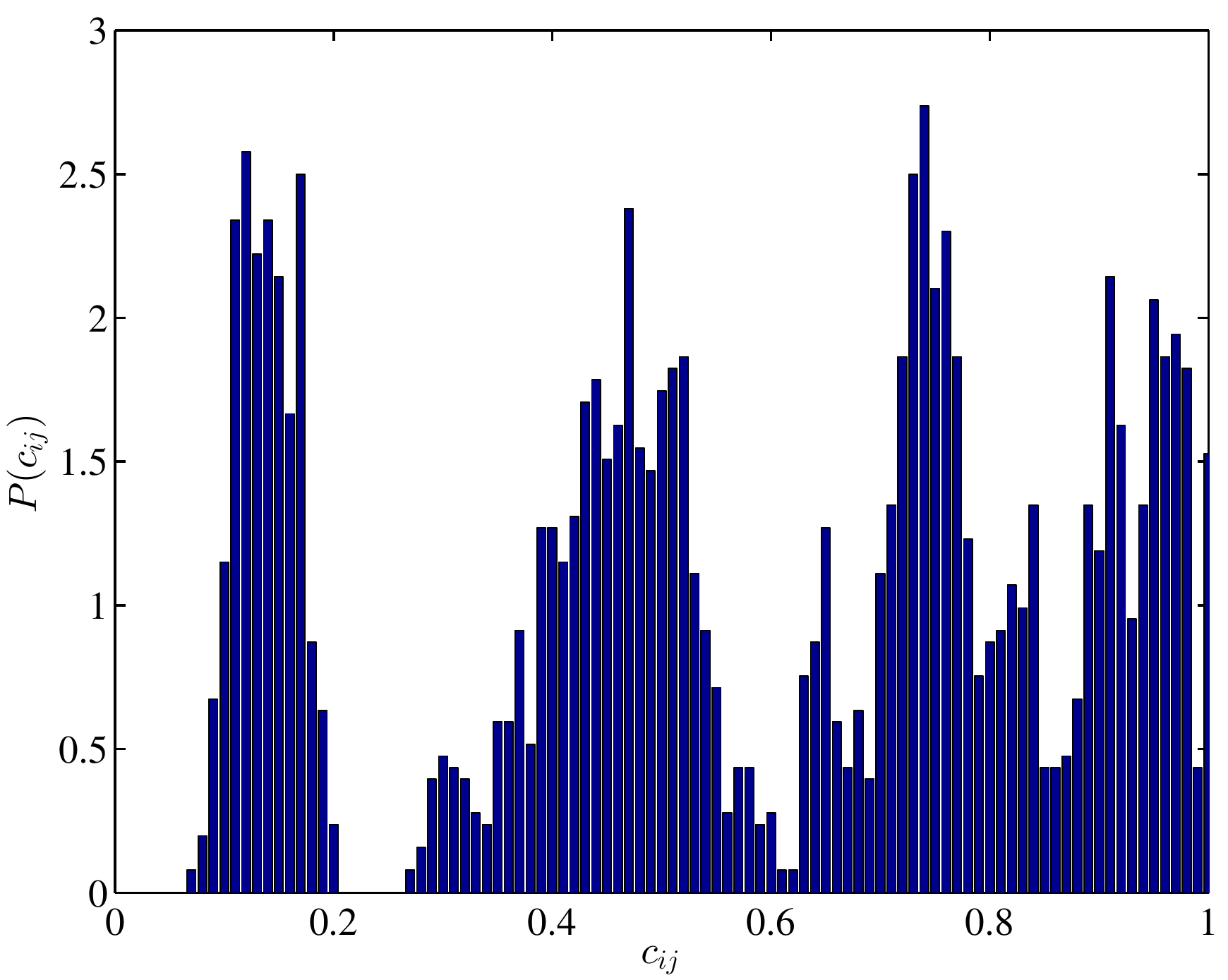}
  \caption{Distribution of correlation coefficients.}\label{Fig:PCA:Oil:PDF:Cij}
\end{figure}

We plot the distribution of correlation coefficients in Fig.~\ref{Fig:PCA:Oil:PDF:Cij}. Four remarked peaks can be easily identified. This feature stems from the fact that the global oil market is less integrated, as shown in Fig.~\ref{Fig:PCA:Oil:Cij}. Such multi-modal patterns in the distribution of correlation coefficients have not been observed in financial markets. Usually, we observe unimodal distributions for financial markets, including the global stock markets \citep{Song-Tumminello-Zhou-Mantegna-2011-PRE} and the US housing market \citep{Meng-Xie-Jiang-Podobnik-Zhou-Stanley-2014-SR} where we observe similar geographical traits.

The right peak corresponds to the largest correlation coefficients between intra-cluster time series, which are along the back diagonal in Fig.~\ref{Fig:PCA:Oil:Cij}(b). The second peak from the right highlights the correlations between inter-cluster time series for clusters 3, 4 and 5. In certain sense, Fig.~\ref{Fig:PCA:Oil:Cij}(b) illustrates that these three clusters can be viewed as a large cluster. The left peak contains correlation coefficients between cluster 1 and clusters 3, 4, 5 and 6. The second peak from the left represents the correlation coefficients between inter-cluster time series other than those in the left peak.

We can see that Fig.~\ref{Fig:PCA:Oil:PDF:Cij} unfolds a picture of the global oil market with a remarked feature of localization.

\section{Eigenvalues and information content}
\label{S1:Eigen}

\subsection{Distribution of eigenvalues}

For the correlation matrix ${\mathbf{C}}$ of each crude oil price series, we can calculate its eigenvalues,
 \begin{equation}
 {\mathbf{C}}={\mathbf{U}}{\mathbf{\Lambda}}{\mathbf{U}}^T,
 \label{Eq:Definition:Eigenvalues}
\end{equation}
where ${\mathbf{U}}$ denotes the eigenvectors, ${\mathbf{\Lambda}}$ is the eigenvalues of the correlation matrix, whose density $f_{c}(\lambda)$ is defined as follows \cite{Laloux-Cizean-Bouchaud-Potters-1999-PRL},
\begin{equation}
 f_{c}(\lambda)=\frac{1}{N}\frac{dn(\lambda)}{d\lambda},
 \label{Eq:PDF:Eigenvalues}
\end{equation}
where $n(\lambda)$ is the number of eigenvalues of ${\mathbf{C}}$ that are less than $\lambda$. If ${\mathbf{M}}$ is a $T$ by $N$ random matrix with zero mean and unit variance, $f_{c}(\lambda)$ is self-averaging. In particular, in the limit $N\rightarrow \infty$, $T\rightarrow \infty$ and $Q=T/N\geq1$ fixed, the probability density function $f_{c}(\lambda)$ of eigenvalues $\lambda$ of the random correlation matrix ${\mathbf{M}}$ has a close form \citep{Edelman-1988-SIAMJmaa,Sengupa-Mitra-1999-PRE,Laloux-Cizean-Bouchaud-Potters-1999-PRL}:
\begin{equation}
 f_{c}(\lambda)=\frac{Q}{2\pi\sigma^2}\frac{\sqrt{(\lambda_{\max}-\lambda)(\lambda-\lambda_{\min})}}{\lambda}
 \label{Eq:RMT:PDF}
\end{equation}
with $\lambda\in[\lambda_{\min}, \lambda_{\max}]$, where $\lambda_{\min}^{\max}$ is given by
\begin{equation}
 \lambda_{\min}^{\max}=\sigma^2(1+1/Q\pm2\sqrt{1/Q})~,
 \label{Eq:RMT:Lambda}
\end{equation}
and $\sigma^2$ is equal to the variance of the elements of ${\mathbf{M}}$ \citep{Sengupa-Mitra-1999-PRE,Laloux-Cizean-Bouchaud-Potters-1999-PRL}. In our case, $Q= 5272/71= 74.3$ and the variance $\sigma^2$ is equal to 1 in our normalized data. If ${\mathbf{C}}$ is a random matrix, the largest eigenvalue $\lambda^{\mathrm{RMT}}_{\max}=1.2456$ and the smallest eigenvalue $\lambda^{\mathrm{RMT}}_{\min}=0.7814$, according to Eq.~(\ref{Eq:RMT:Lambda}).

\begin{figure}
  \centering
  \includegraphics[width=8cm]{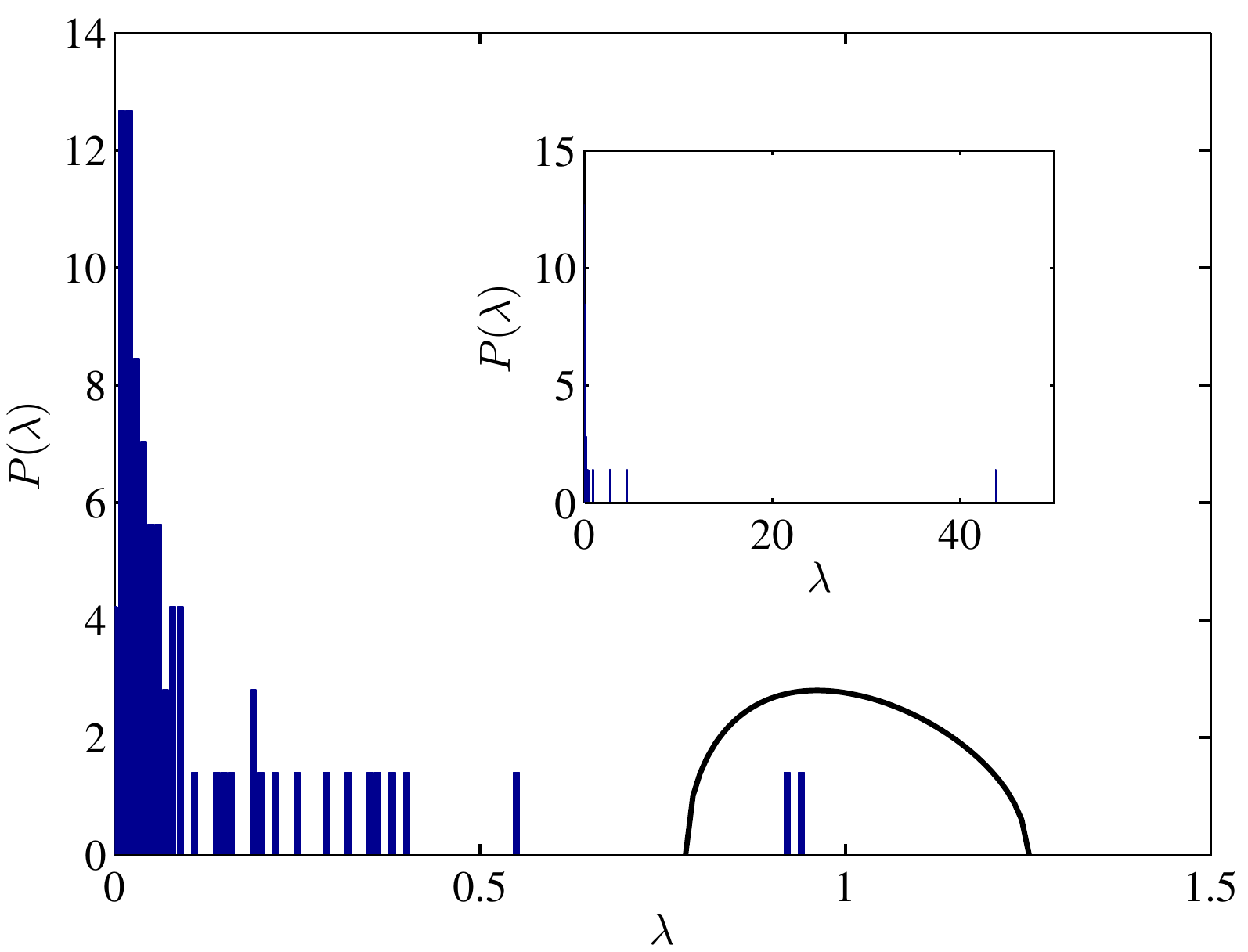}
  \caption{Empirical distribution of eigenvalues of the correlation matrix ${\mathbf{C}}$ with the large eigenvalues in the inset. The solid line is the theoretical distribution expressed in Eq.~(\ref{Eq:RMT:PDF}) from the random matrix theory predictions.}\label{Fig:PCA:Oil:PDF:Eigenvalues}
\end{figure}

We compute the $N$ eigenvalues of the correlation matrix ${\mathbf{C}}$, $\lambda_{\max}=\lambda_{1}>\lambda_{2}>\cdots>\lambda_{70}>\lambda_{71}=\lambda_{\min}$. We find that the largest eigenvalue $\lambda_{\max}=43.76$ and the smallest eigenvalue $\lambda_{\min}=0.00073$. Figure \ref{Fig:PCA:Oil:PDF:Eigenvalues} illustrates the empirical distribution of these eigenvalues and the theoretical distribution based on the random matrix theory. There are only two eigenvalues $\lambda_6$ and $\lambda_7$ falling in the theoretical curve. The five largest eigenvalues are greater than $\lambda^{\mathrm{RMT}}_{\max}$ and the other 64 eigenvalues are smaller than $\lambda^{\mathrm{RMT}}_{\min}$. The deviation of the empirical distribution from the theoretical prediction implies that the correlation matrix $\mathbf{C}$ is not a random matrix and there is economic information embedded in the deviating eigenvalues.

\subsection{Eigenportfolios}

To uncover the information contents in the deviating eigenvalues, we construct eigenportfolios for each eigenvalue. For $\lambda_k$, we have
\begin{equation}
 R_k(t)=\frac{{\mathbf{u}}_k^T\cdot{\mathbf{r}}}{\sum_{i=1}^{N}u_{i,k}}=\frac{1}{\sum_{i=1}^{N}u_{i,k}}  \sum_{i=1}^{N} u_{i,k}r_i(t),
 \label{Eq:EigDefinition:Definition}
\end{equation}
where ${\mathbf{u}}_k^T\cdot{\mathbf{r}}$ is the projection of the normalized return vector ${\mathbf{r}}$ on the $k$-th eigenvector ${\mathbf{u}}_k$.

\begin{figure}[tbh]
  \centering
  \includegraphics[width=4cm,height=3.2cm]{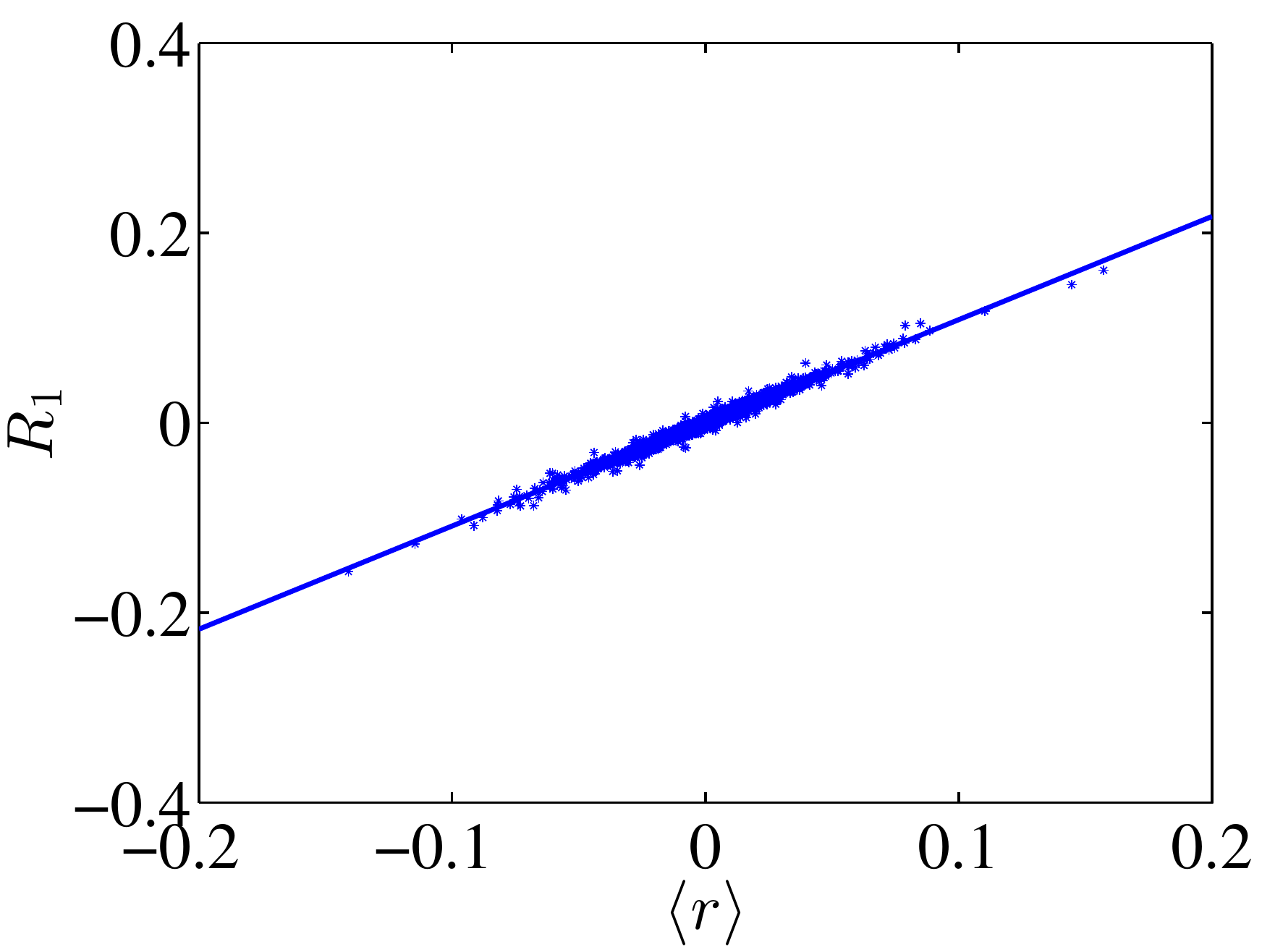}
  \includegraphics[width=4cm,height=3.2cm]{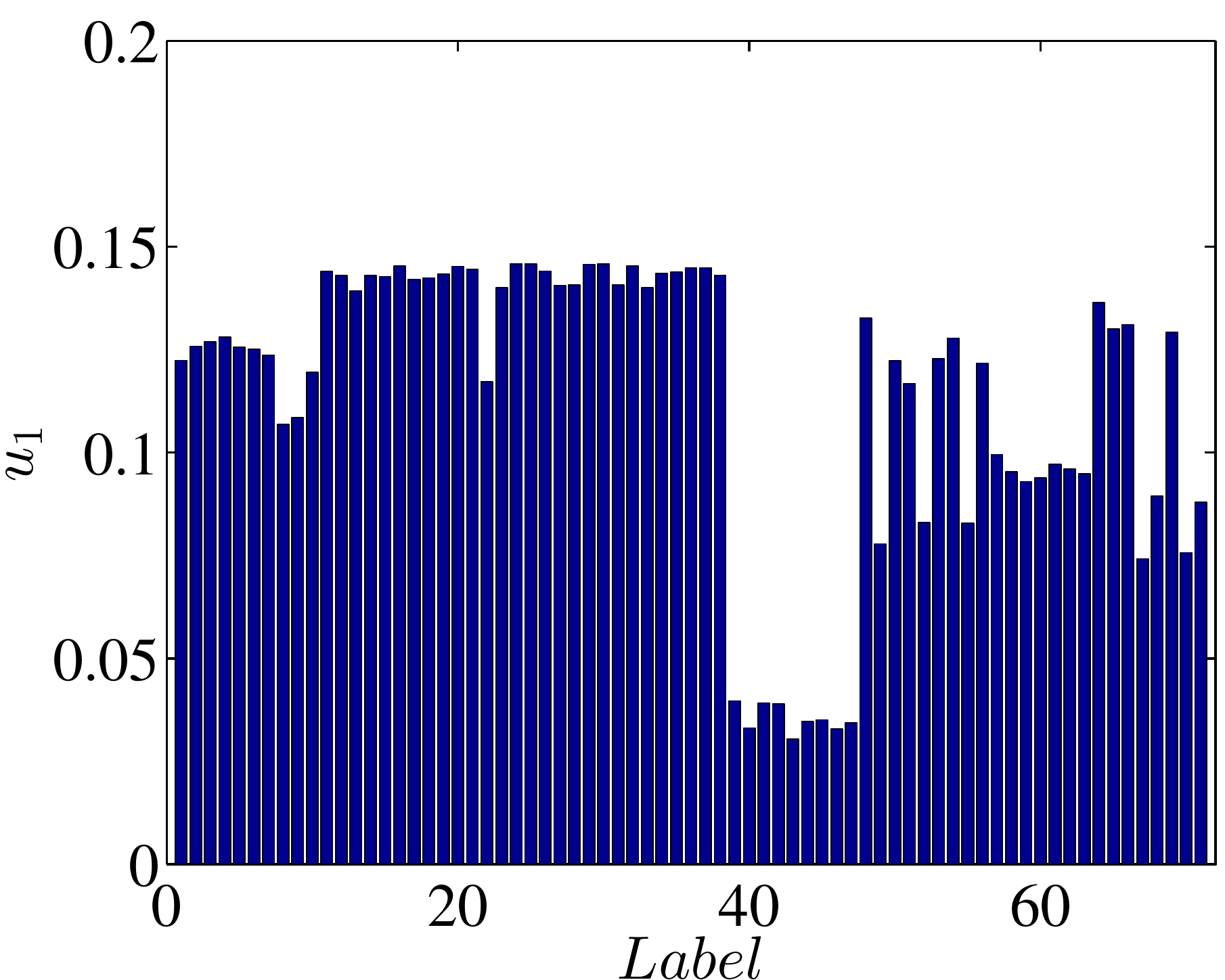}\\
  \includegraphics[width=4cm,height=3.2cm]{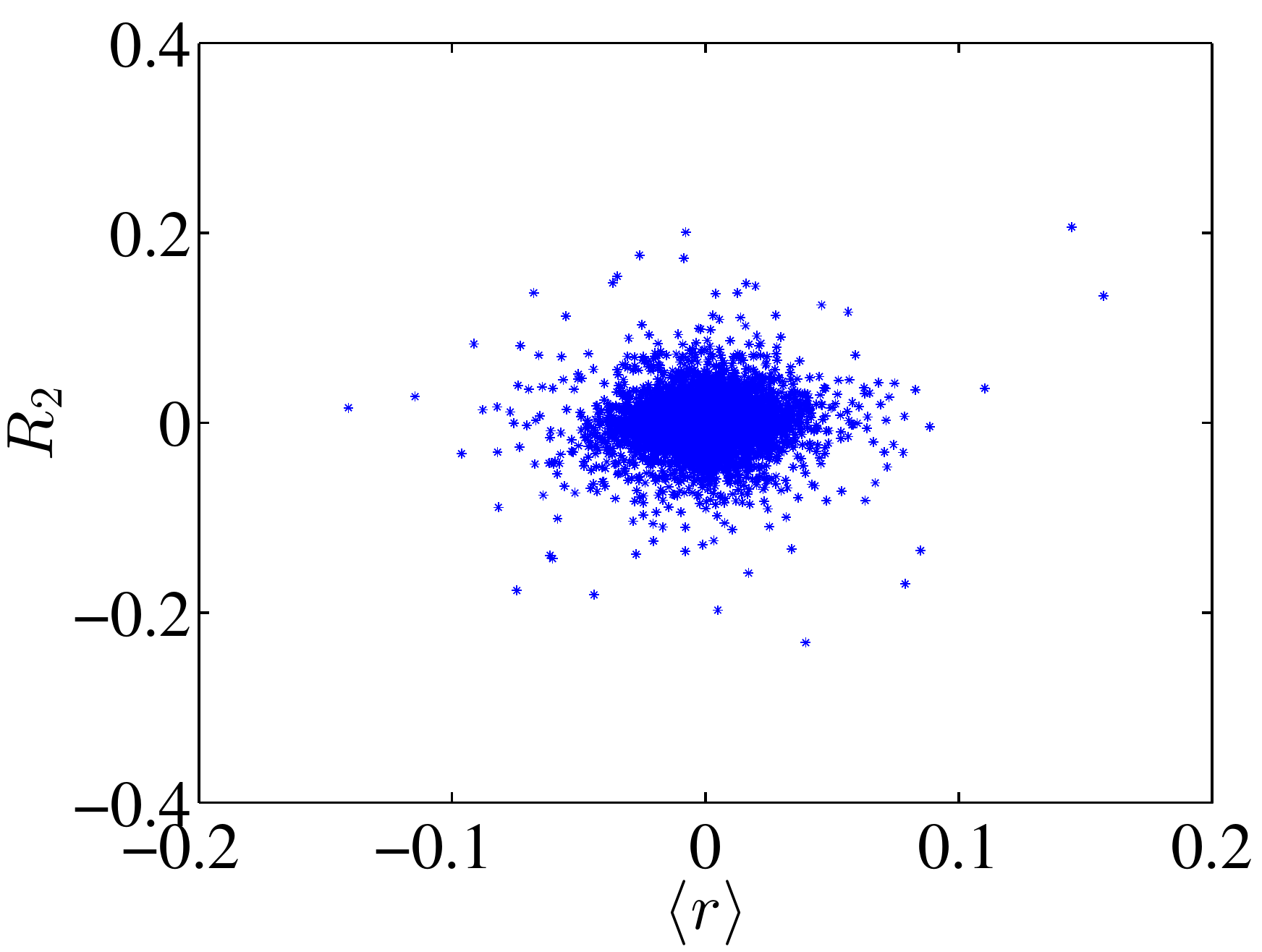}
  \includegraphics[width=4cm,height=3.2cm]{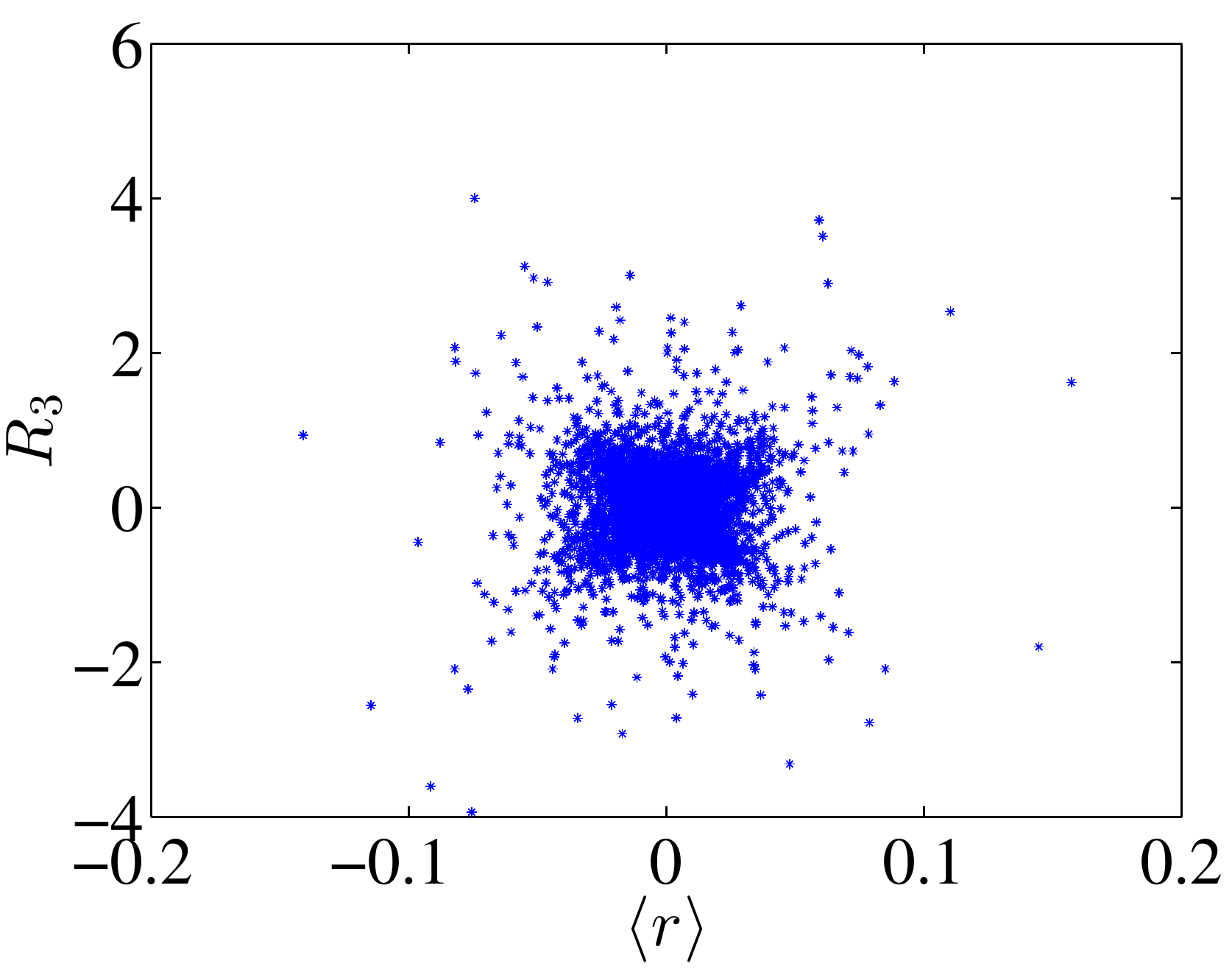}\\
  \includegraphics[width=4cm,height=3.2cm]{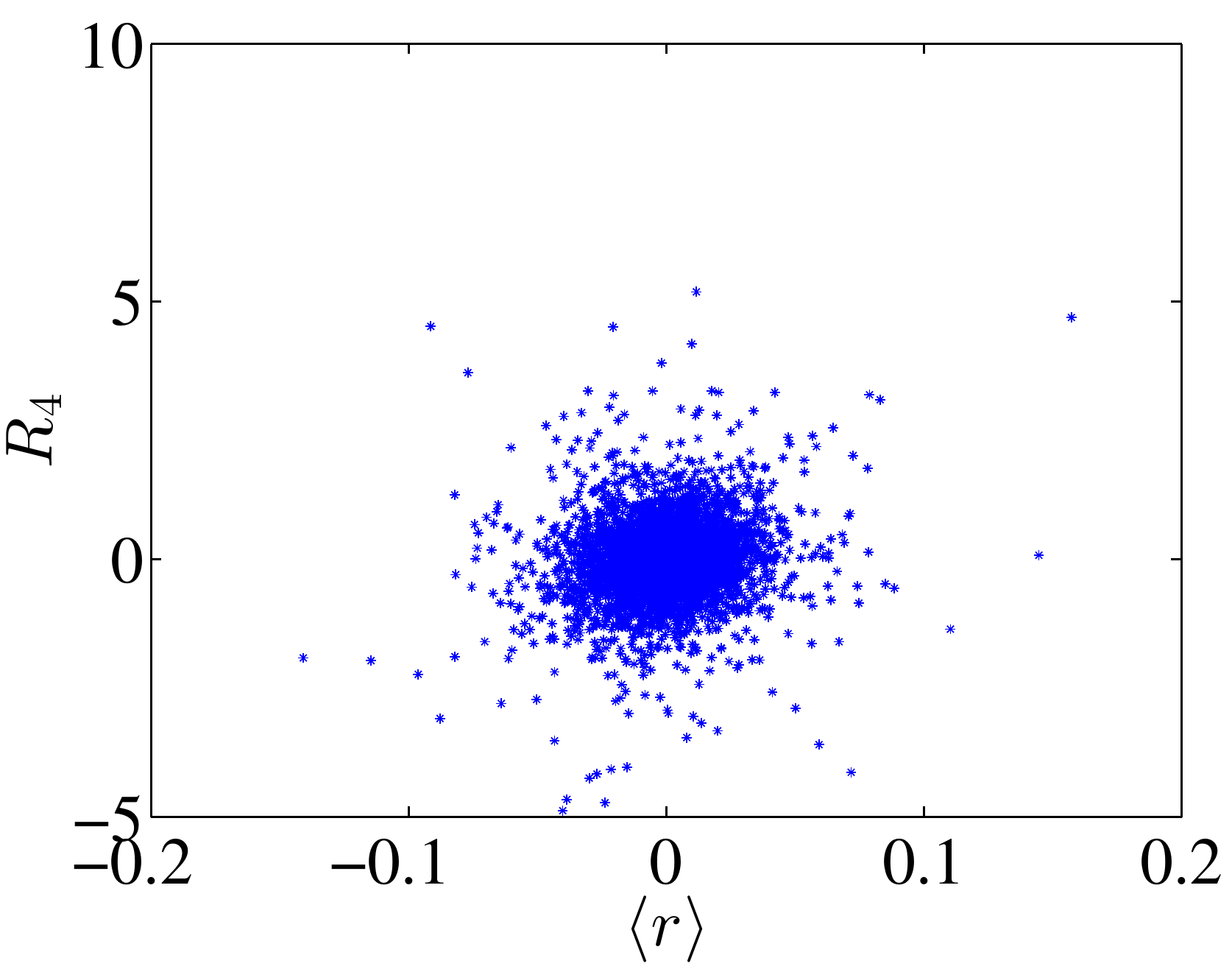}
  \includegraphics[width=4cm,height=3.2cm]{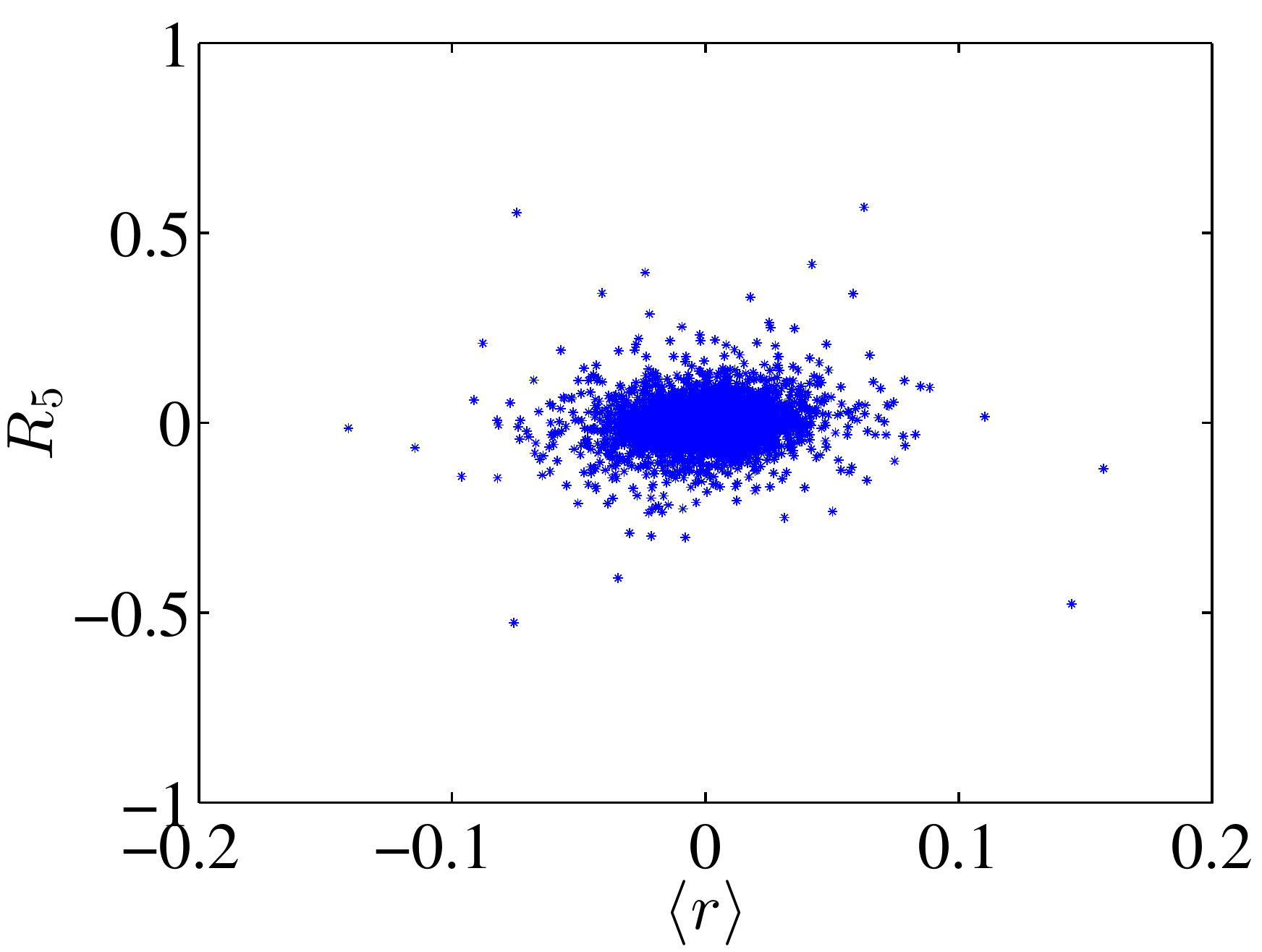}
  \vskip  -9.55cm   \hskip   -6.32cm {\textsf{(a)}}
  \vskip  -0.40cm   \hskip    1.70cm {\textsf{(b)}}
  \vskip   2.80cm   \hskip   -6.32cm {\textsf{(c)}}
  \vskip  -0.40cm   \hskip    1.70cm {\textsf{(d)}}
  \vskip   2.80cm   \hskip   -6.22cm {\textsf{(e)}}
  \vskip  -0.40cm   \hskip    1.8cm {\textsf{(f)}}
  \vskip 2.5cm
  \caption{Testing collective market effect. (a) Relationship between the mean returns $\langle{r}\rangle$ and the eigenportfolio $R_1$. (b) Components of the eigenvector ${\mathbf{u}}_1$ of the largest eigenvalue $\lambda_1$. (c) Relationship between the mean returns $\langle{r}\rangle$ and the eigenportfolio $R_2$. (d) Relationship between the mean returns $\langle{r}\rangle$ and the eigenportfolio $R_3$. (e) Relationship between the mean returns $\langle{r}\rangle$ and the eigenportfolio $R_4$. (f) Relationship between the mean returns $\langle{r}\rangle$ and the eigenportfolio $R_5$. }\label{Fig:PCA:Oil:Rk:r}
\end{figure}

Figure \ref{Fig:PCA:Oil:Rk:r}(a) illustrates the relationship between the eigenportfolio $R_1$ and the mean return $\langle{r}\rangle$. There is an evident linear relationship with a very high R-square of 0.97. Hence, the largest eigenvalue reflects a common factor that drives the global crude oil market. Because the largest eigenvalue explains $\lambda_1/N=61.63\%$ of the variation of the price fluctuations, this common factor implies a collective effect of the whole market mode. In addition, Fig.~\ref{Fig:PCA:Oil:Rk:r}(b) shows that all the components of the first eigenvector have the same sign and are not randomly distributed. However, no evident relationship between the mean return and the eigenportfolios is observed for other eigenvalues, which means that other deviating eigenvalues do not bear any market-wide effects. These observations are also reported for stock markets \citep{Plerou-Gopikrishnan-Rosenow-Amaral-Guhr-Stanley-2002-PRE}.

Because the largest eigenvalue reflects the global movement of the crude oil markets, we can construct an index for the global crude oil market based on the eigenportfolio of the largest eigenvalue:
\begin{equation}
  I(t) = \langle{P(0)}\rangle\exp\left[\sum_{t=1}^TR_1(t)\right]
  \label{Eq:PCA:Oil:Index}
\end{equation}
where $\langle{P(0)}\rangle=19.4704$ is the average price on October 2, 1992. We compare the constructed index with the average price in Fig.~\ref{Fig:PCA:Oil:Mean:Price:Return}(a). It is found that the evolution of the index is very similar to the average price as expected. Moreover, the index performs better than the average price under the buy and hold strategy because the index is always greater than the average price.

\subsection{Other largest eigenvalues and the partitioning effect}

\begin{figure}
  \centering
  \includegraphics[width=4cm]{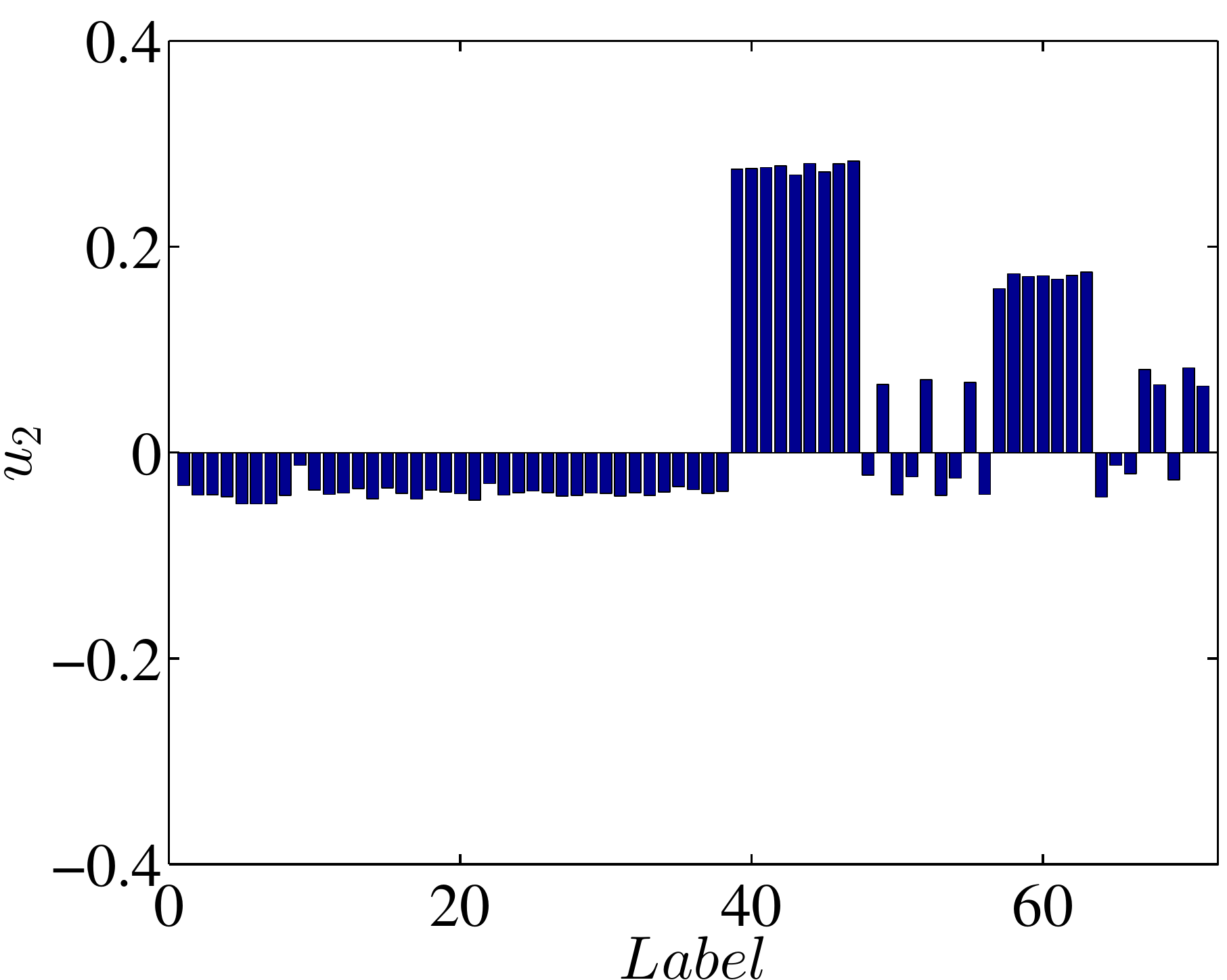}
  \includegraphics[width=4cm]{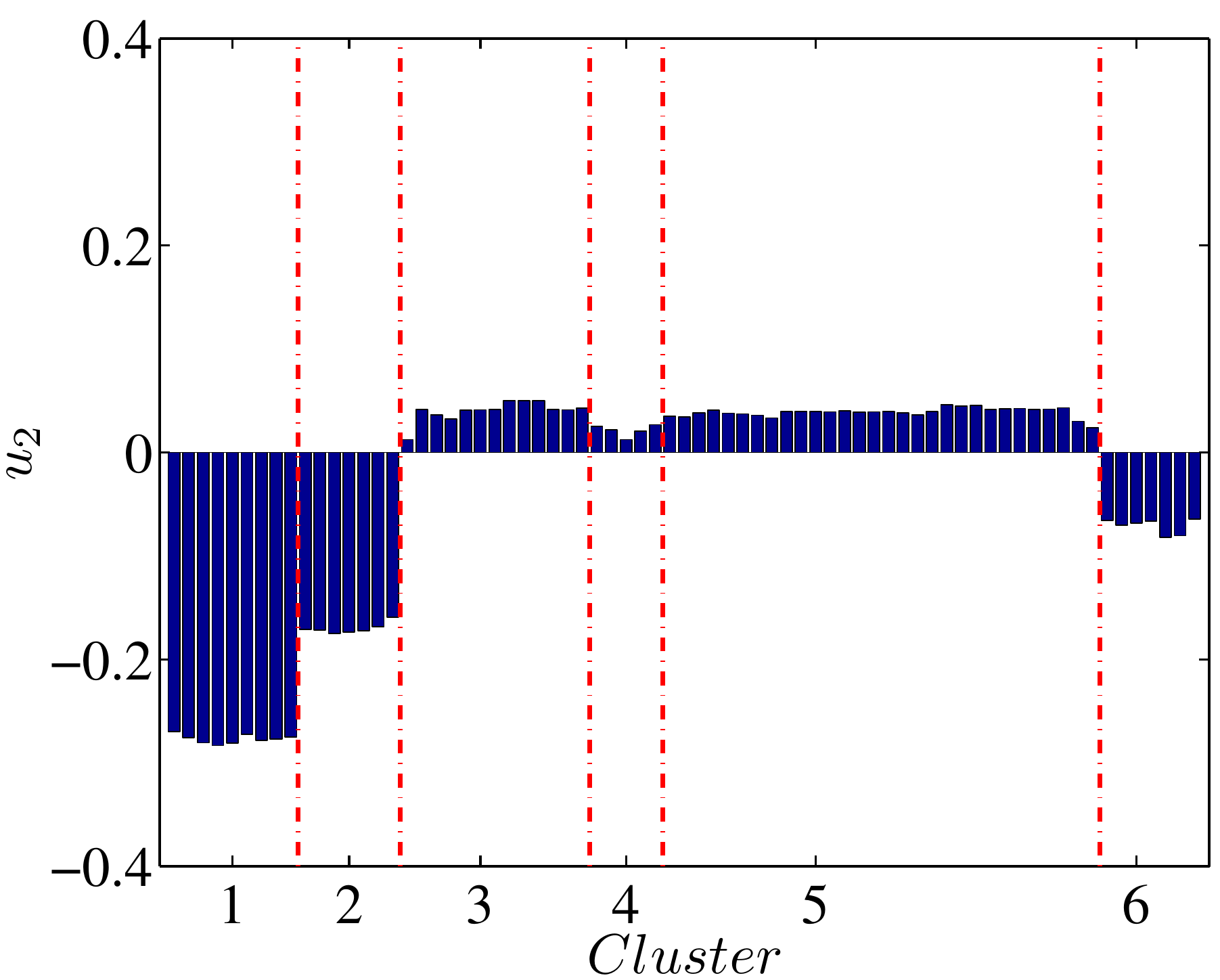}\\
  \includegraphics[width=4cm]{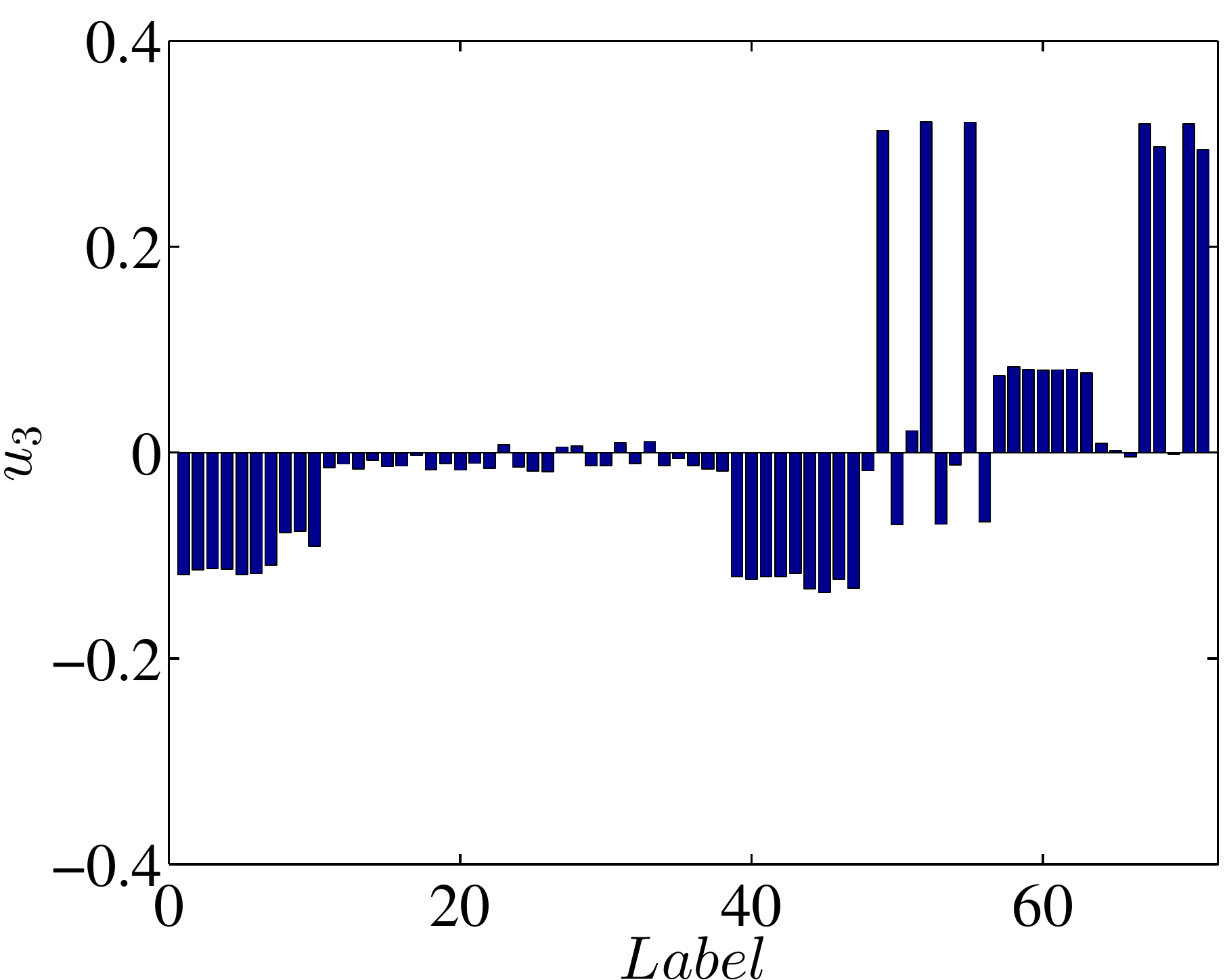}
  \includegraphics[width=4cm]{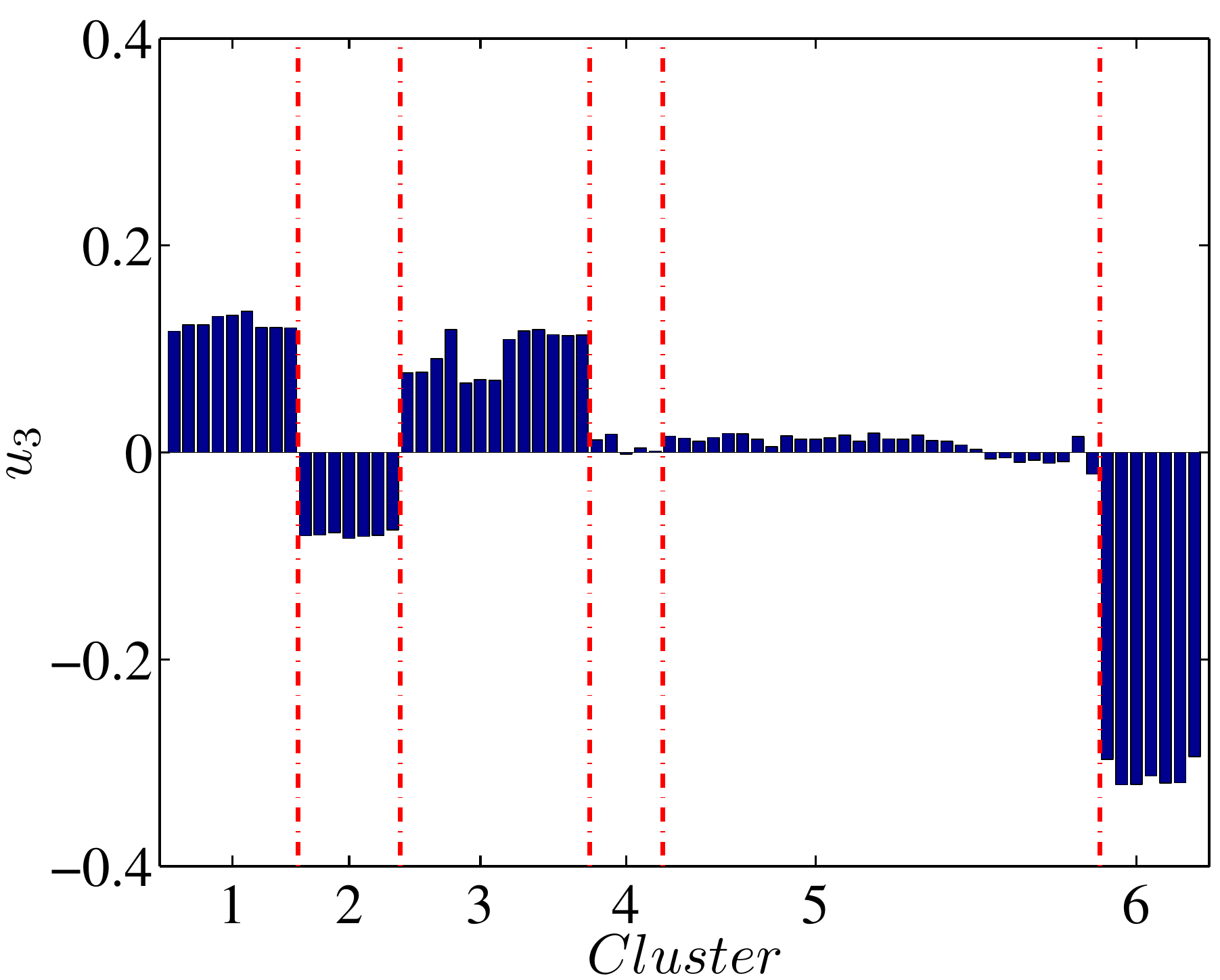}\\
  \includegraphics[width=4cm]{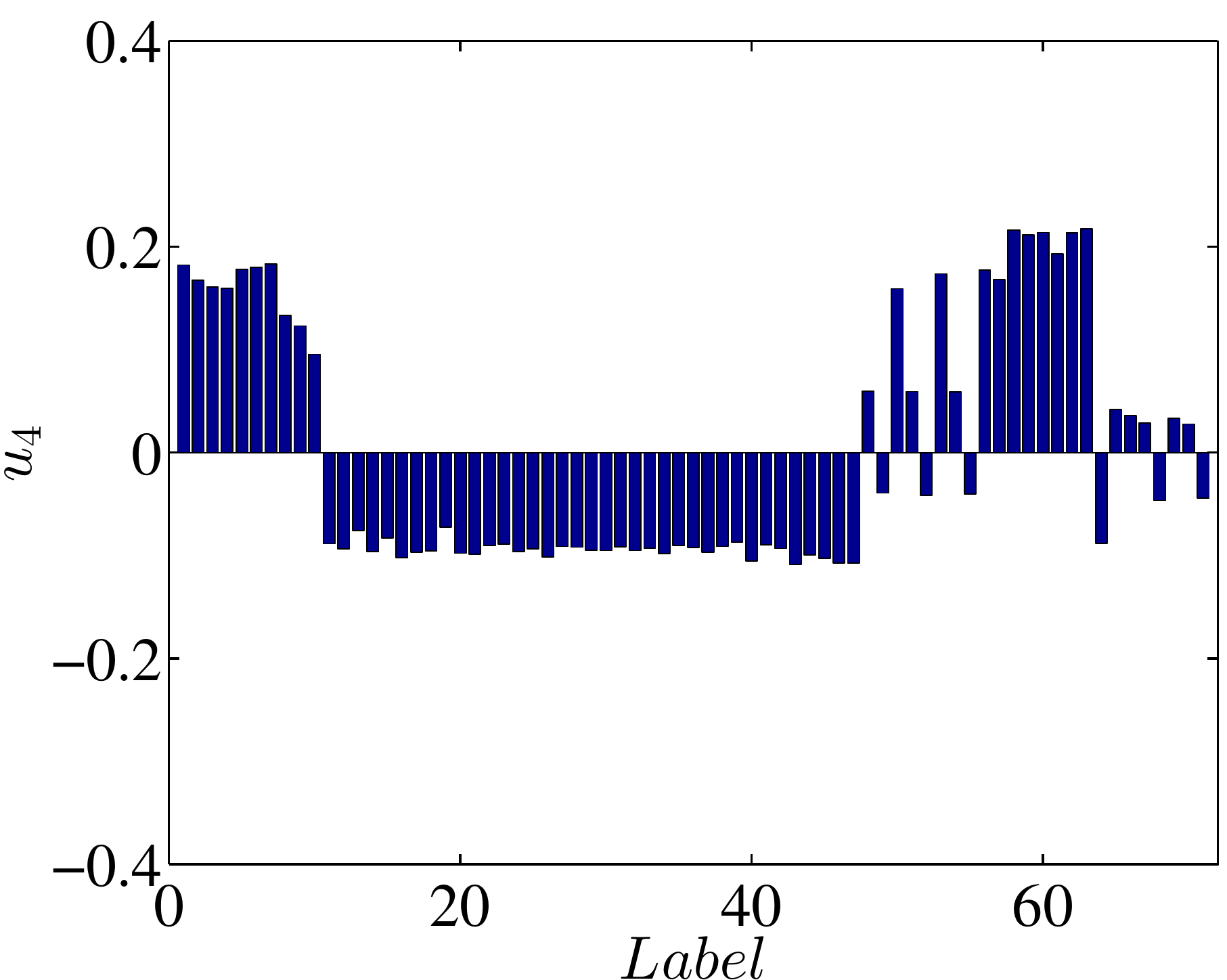}
  \includegraphics[width=4cm]{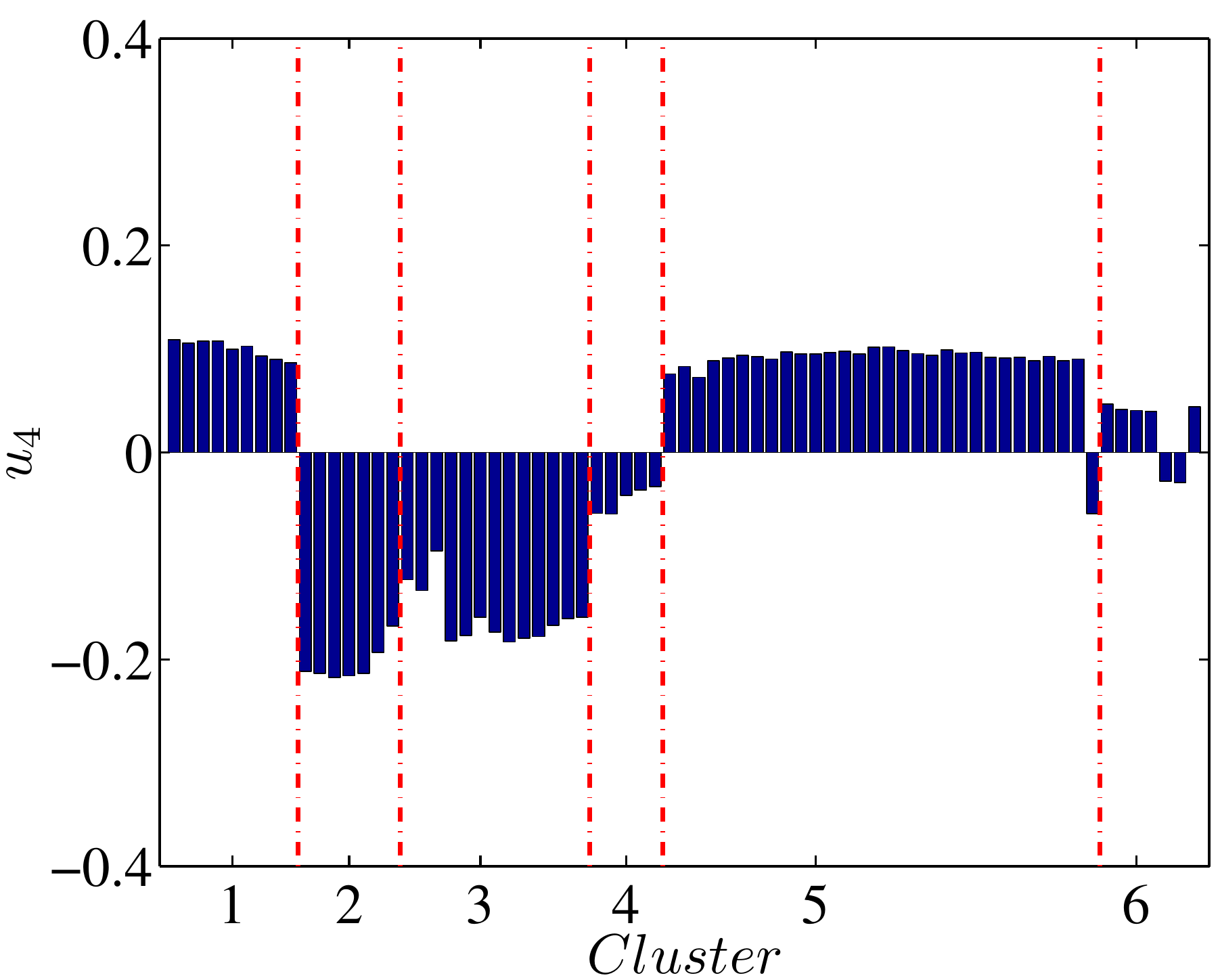}\\
  \includegraphics[width=4cm]{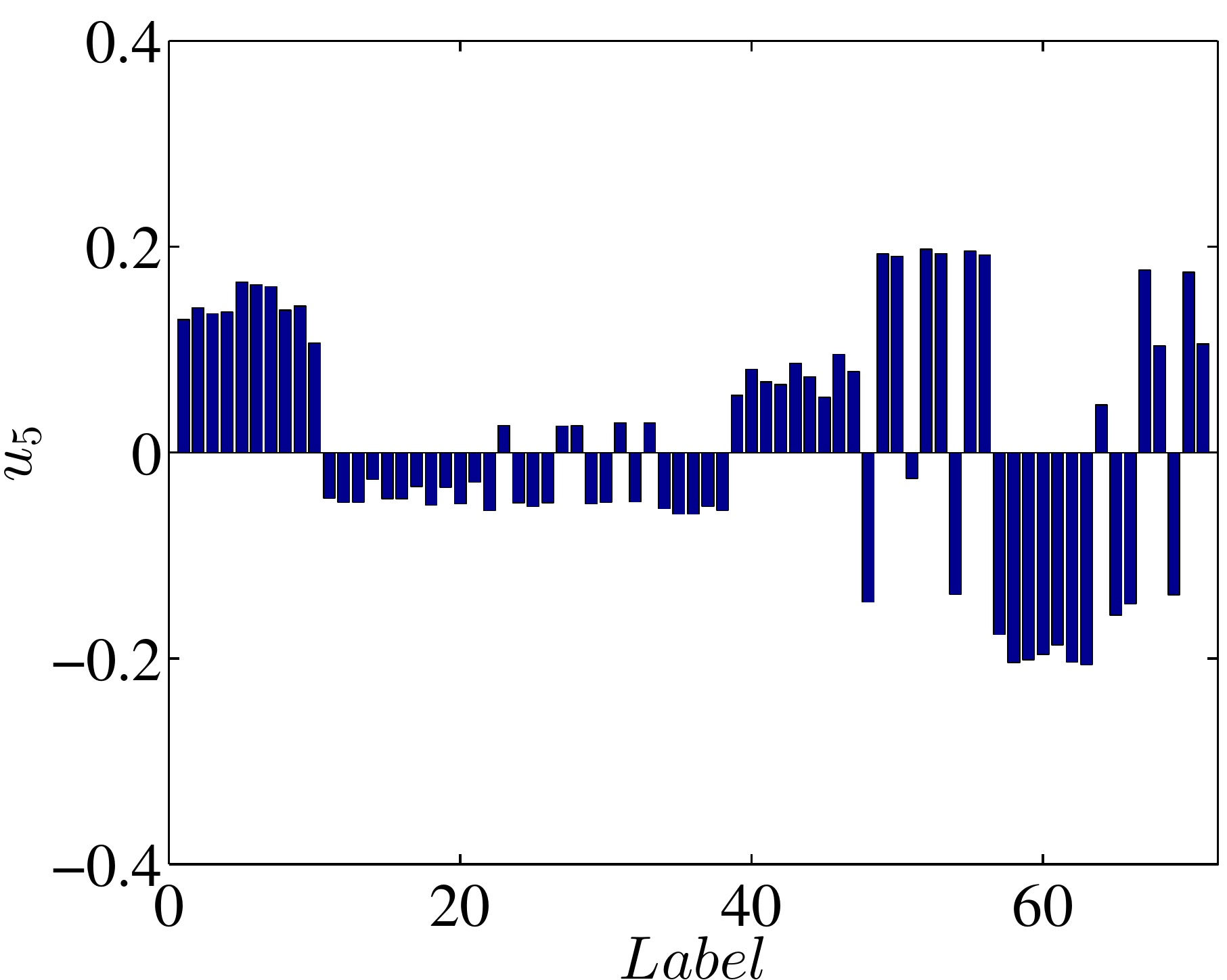}
  \includegraphics[width=4cm]{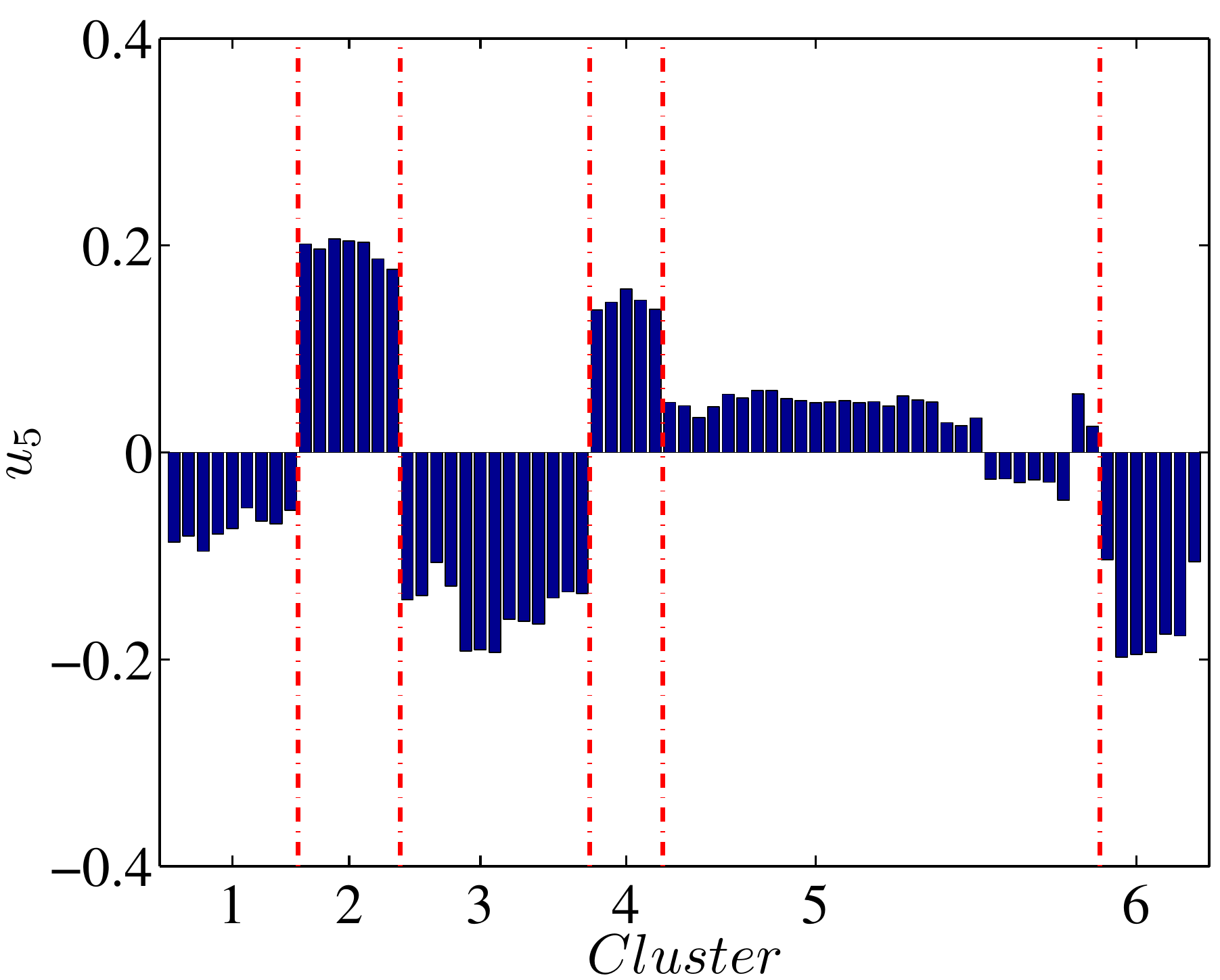}
  \vskip  -12.8cm   \hskip   -6.32cm {\textsf{(a)}}
  \vskip  -0.40cm   \hskip    1.70cm {\textsf{(b)}}
  \vskip   2.80cm   \hskip   -6.32cm {\textsf{(c)}}
  \vskip  -0.40cm   \hskip    1.70cm {\textsf{(d)}}
  \vskip   2.80cm   \hskip   -6.22cm {\textsf{(e)}}
  \vskip  -0.40cm   \hskip    1.70cm {\textsf{(f)}}
  \vskip   2.80cm   \hskip   -6.22cm {\textsf{(g)}}
  \vskip  -0.40cm   \hskip    1.70cm {\textsf{(h)}}
  \vskip 2.5cm
  \caption{\label{Fig:PCA:Oil:Largest:eigenvectors} (Color online) Partitioning effect of the eigenvectors ${\mathbf{u}}_2$ (a, b), ${\mathbf{u}}_3$ (c, d), ${\mathbf{u}}_4$ (e, f), and ${\mathbf{u}}_5$ (g, h) associated with the four deviating eigenvalues $\lambda_2$ to $\lambda_5$. Left panel: Raw eigenvectors; Right panel: Reordered eigenvectors based on the six clusters.}
\end{figure}

For stock markets, other deviating eigenvalues that are greater than $\lambda^{\mathrm{RMT}}_{\max}$ contain partitioning information of industrial sectors \citep{Plerou-Gopikrishnan-Rosenow-Amaral-Guhr-Stanley-2002-PRE}. Following this idea, we show the four eigenvectors in the left panel of Fig.~\ref{Fig:PCA:Oil:Largest:eigenvectors}. We can already see clusters of components with similar heights. These clusters of components correspond to the clusters identified in Fig.~\ref{Fig:PCA:Oil:Cij}. To further illustrate this point, we plot in the right panel of Fig.~\ref{Fig:PCA:Oil:Largest:eigenvectors} the reordered eigenvectors based on the six clusters.

\subsection{Smallest eigenvalues and highly correlated pairs}

Having unveiled the latent information carried by the deviating eigenvalues greater than $\lambda^{\mathrm{RMT}}_{\max}$ and their corresponding eigenvectors, we now turn to investigate the hidden information in the smallest eigenvalues. According to \cite{Plerou-Gopikrishnan-Rosenow-Amaral-Guhr-Stanley-2002-PRE}, the smallest eigenvalues and their corresponding eigenvectors stand for the high-correlation time series pairs of stocks. We find similar information in the global crude oil market.

 \begin{figure}[t]
  \centering
  \includegraphics[width=4cm]{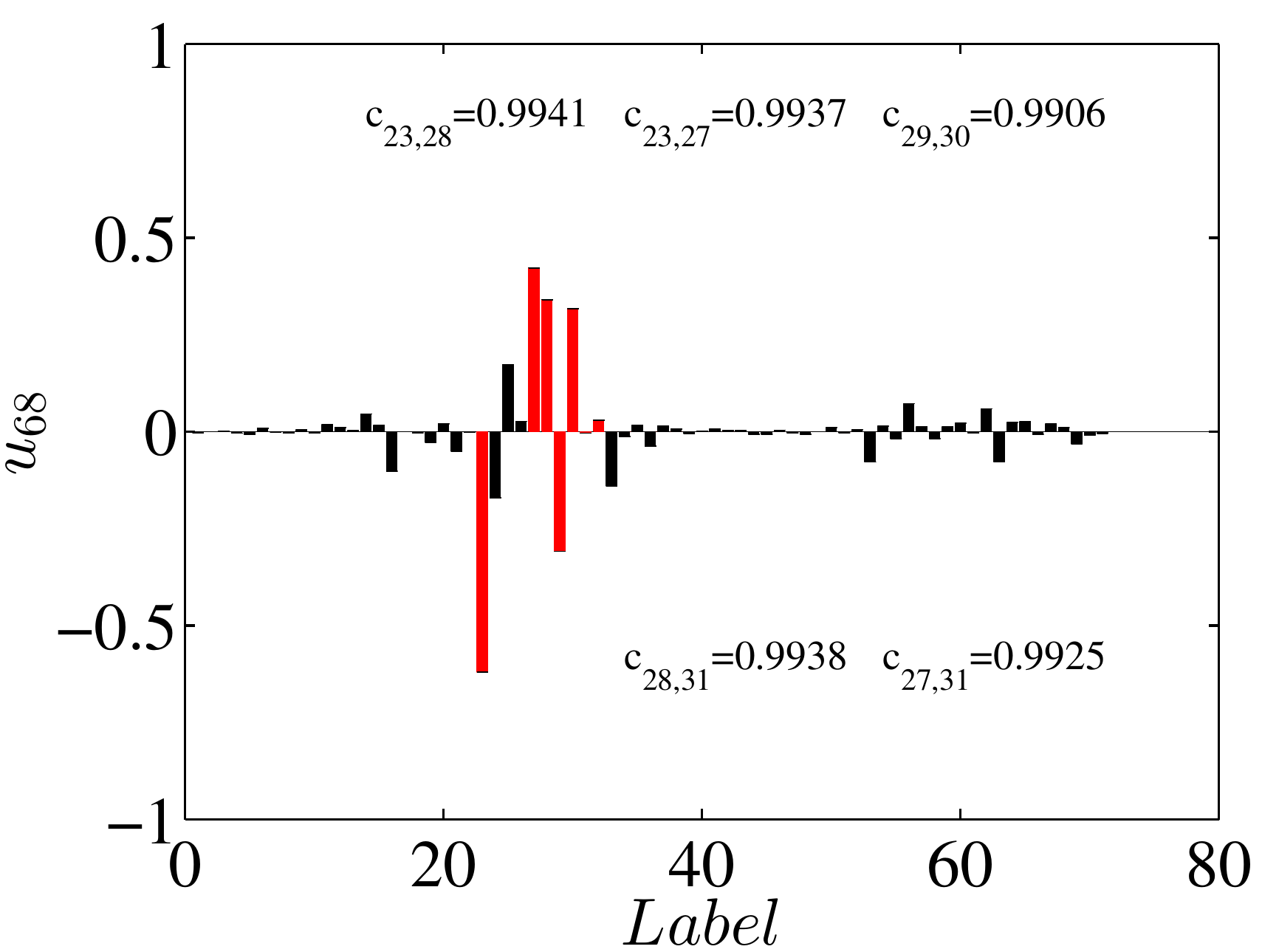}
  \includegraphics[width=4cm]{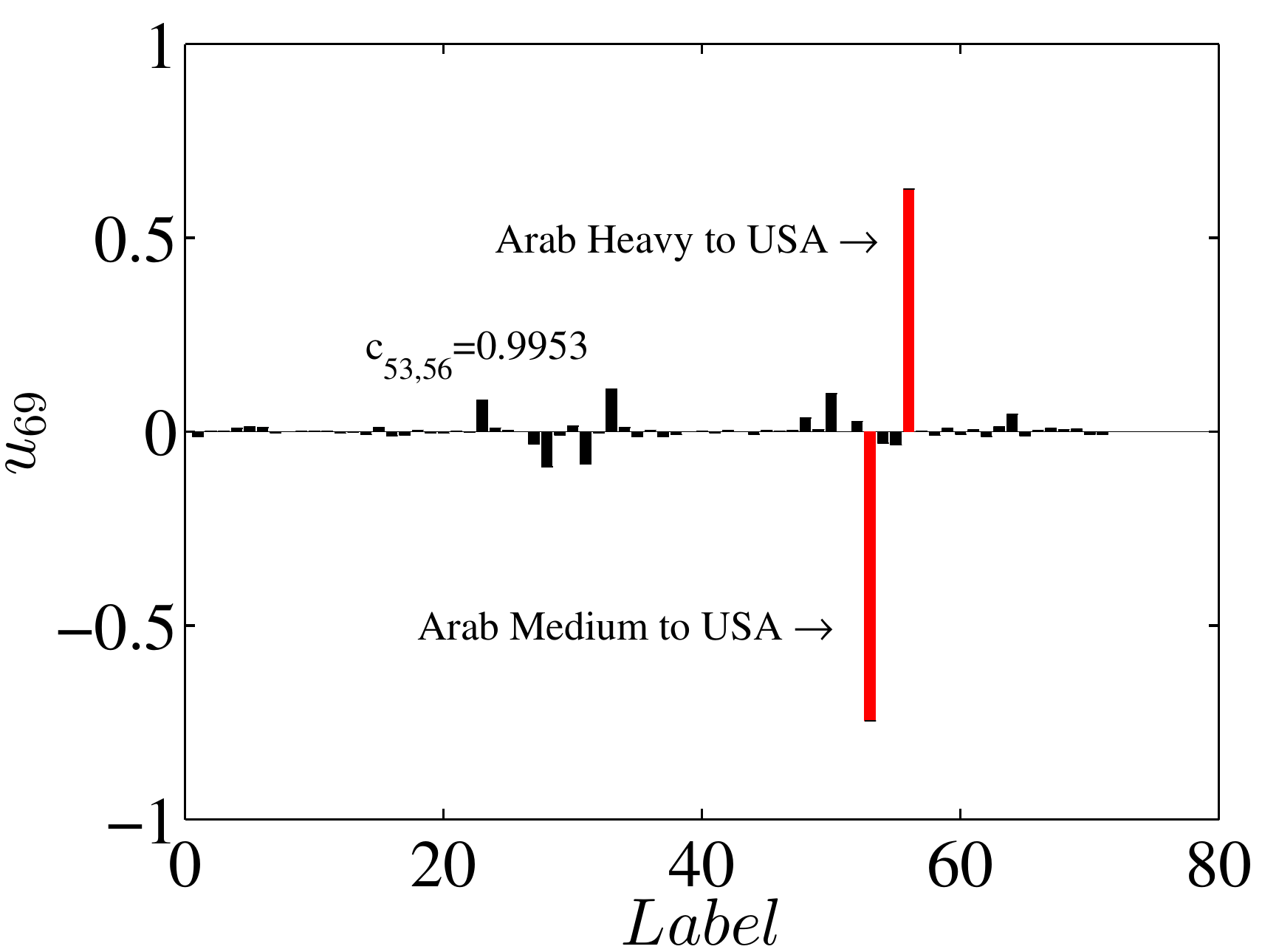}\\
  \includegraphics[width=4cm]{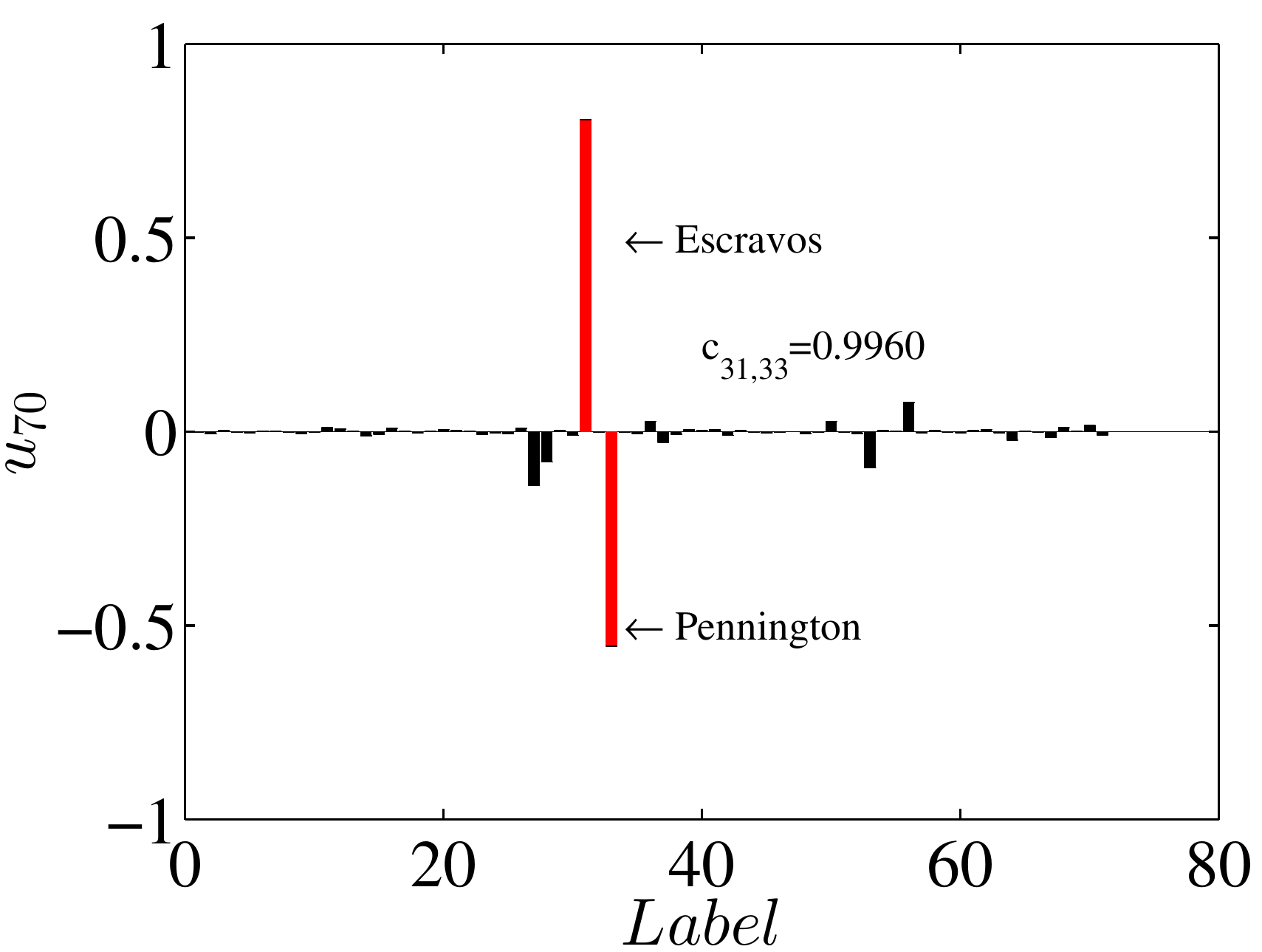}
  \includegraphics[width=4cm]{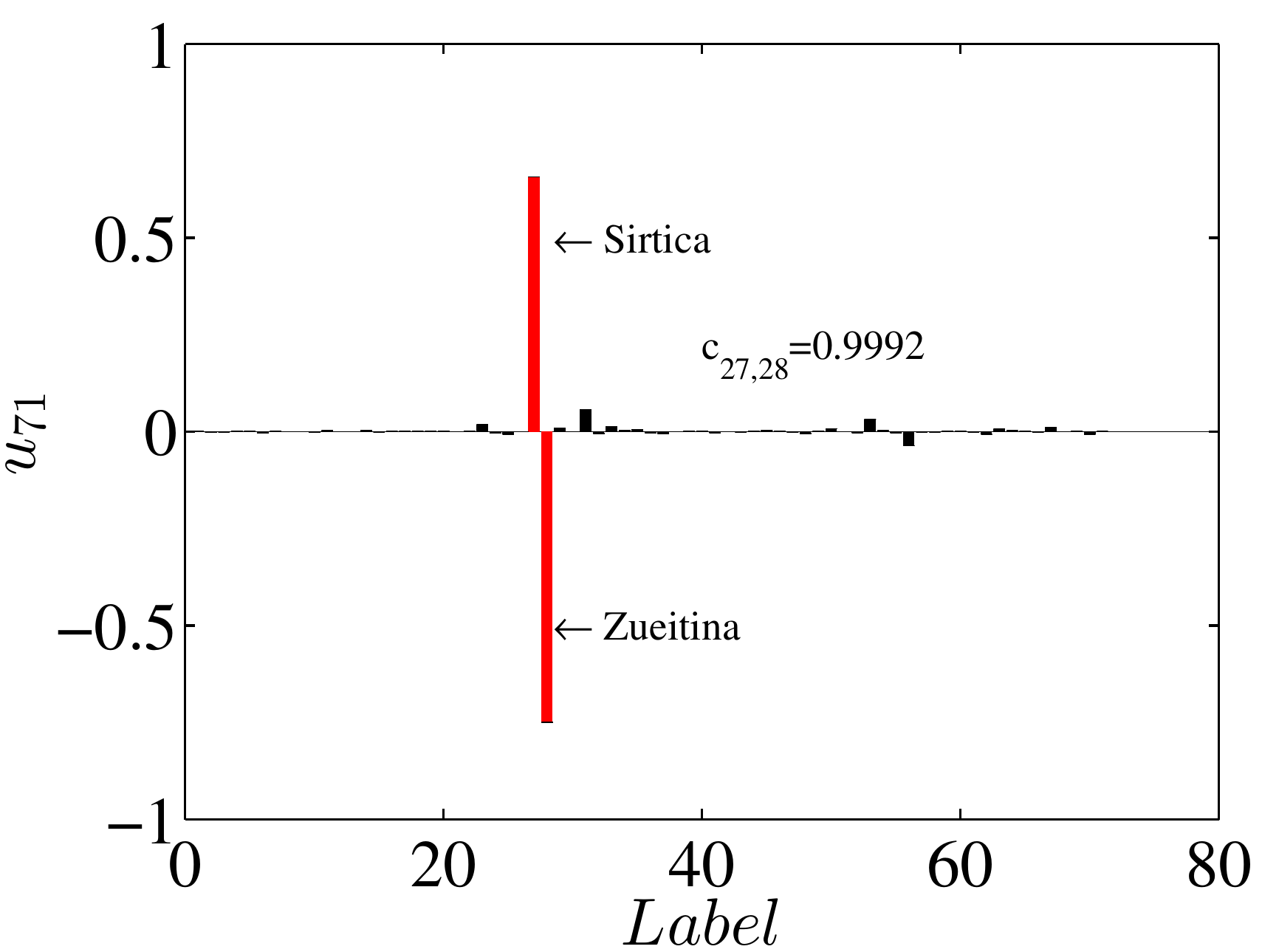}
  \caption{\label{Fig:PCA:Oil:Smallest:eigenvectors} (Color online) Eigenvectors ${\mathbf{u}}_{68}$, ${\mathbf{u}}_{69}$, ${\mathbf{u}}_{70}$, and ${\mathbf{u}}_{71}$ of the four smallest eigenvalues $\lambda_{68}$ to $\lambda_{71}$.}
\end{figure}

Figure \ref{Fig:PCA:Oil:Smallest:eigenvectors} shows the eigenvectors of the four smallest eigenvectors. In each plot of ${\mathbf{u}}_{69}$, ${\mathbf{u}}_{70}$ and ${\mathbf{u}}_{71}$, we observe a pair of components with opposite signs and significant large magnitudes. We locate the price label of each component and calculate the corresponding correlation coefficient between each pair of return time series. The two time series for ${\mathbf{u}}_{71}$ are Sirtica and Zueitina, whose correlation coefficient $c_{27,28}=0.9992$ is the highest in all correlation coefficients. The pair for ${\mathbf{u}}_{70}$ are Escravos and Pennington,  whose correlation coefficient $c_{31,33}=0.9960$ is the second largest among all $c_{ij}$'s and the pair for eigenvector ${\mathbf{u}}_{69}$ are Arab Medium to USA and Arab Heavy to USA with $c_{53,56}=0.9953$, which is third largest.

The situation for ${\mathbf{u}}_{68}$ is less clear and complicated, because there are several ``bars'' that outstand. We can nevertheless identify several pairs for the longest bars: Amna and Zueitina with $c_{23,28}=0.9941$ being the fourth largest correlation coefficient, Amna and Sirtica with $c_{23,27}=0.9937$ being the sixth largest correlation coefficient, and Bonny Light and Brass River with $c_{29,30}=0.9906$ being the eighth largest coefficient. We also find that the fifth largest is $c_{28,31}=0.9938$ between Zueitina and Escravos and the seventh largest is $c_{27,31}=0.9925$ between Sirtica and Escravos, in which the 31st component of ${\mathbf{u}}_{68}$ is small.

Of these time series in the pairs, seven series (23, 27, 28, 29, 30, 31 and 33) belong to Cluster 5, while two (53 and 56) belong to Cluster 3. This finding is reasonable because time series in the same cluster usually have large correlation coefficient as shown in Fig.~\ref{Fig:PCA:Oil:Cij}(b).

\section{Conclusions}
\label{S1:Conclusion}

We have investigated the correlation structure of the global crude oil market including 71 time series of daily oil prices all over the world by performing principal component analysis. Several empirical statistical properties of the global oil market have been unveiled.

We found that the distribution of correlation coefficients between pairs of daily returns is not unimodal, but multi-modal. We also identified six clusters of price time series from the correlation matrix, which have evident geographical traits. The multi-modality of the correlation coefficient distribution is caused by the geographical traits of the price time series and the peaks correspond well to the distinct intra-cluster and inter-cluster correlations.

We found that most eigenvalues of the correlation matrix locate outside the prediction of the random matrix theory, where the five largest eigenvalues are greater than the theoretical maximal eigenvalue $\lambda^{\mathrm{RMT}}_{\max}$ and 64 small eigenvalues are less than the theoretical minimal eigenvalue $\lambda^{\mathrm{RMT}}_{\min}$. We showed these deviating eigenvalues and their corresponding eigenvectors embed certain economic information. Specifically, the largest eigenvalue reflects a collective effect of the global market, other four largest eigenvalues bear geographical information that is capable of identifying the six clusters, and the smallest eigenvalues usually have very large components in the eigenvectors which corresponds to pairs of time series with the largest correlation coefficients in the correlation matrix.

We constructed eigenportfolios based on the eigenvectors. It is found that the returns of the eigenportfolio for the largest eigenvalue correlate strongly to the average returns of the oil prices, while other eigenportfolios do not exhibit significant correlations with the average returns. Inspired by this feature, we proposed to construct an index for the global crude oil market based on the eigenportfolio of the largest eigenvalue. This index evolves very similarly to the average price time series and outperforms the latter under the buy-and-hold strategy.

\section*{Acknowledgements}

This work was supported by National Natural Science Foundation of China (Grant No. 11075054), Shanghai (Follow-up) Rising Star Program (Grant No. 11QH1400800), and Fundamental Research Funds for the Central Universities.

%
%\bibliographystyle{elsarticle-harv}
%\bibliography{E:/Papers/Auxiliary/Bibliography}

\clearpage
\newpage

\begin{center}
\begin{longtable}{cccccccccccccccccc}
  \caption{Summary statistics of the daily returns of 71 crude oil price time series all over the world.}
  \label{TB:Oil:Datasets}\\
  \hline
  \hline
  Label & Time series & Length & Max & Min & Mean($\times 10^4$) & s.t.d. & Skewness & Kurtosis & Cluster  \\
  \hline
1&Aloska North Slope&5518&0.372&-0.277&3.739 &0.027&0.190&17.142&3\\
2&Eugene Island&5417&0.201&-0.183&3.075 &0.025&0.021&7.741&3\\
3&Heavy Louiana Sweet&6255&0.178&-0.191&3.506 &0.025&-0.156&7.893&3\\
4&Light Louiana Sweet&7371&0.178&-0.183&2.281 &0.024&-0.195&8.811&3\\
5&WTI Cushing&7383&0.213&-0.173&2.010 &0.024&-0.179&9.305&3\\
6&WTI Midland&6255&0.205&-0.192&2.929 &0.025&-0.144&8.443&3\\
7&West Texas Sour&6255&0.191&-0.163&3.110 &0.027&-0.005&7.151&3\\
8&Isthmus to USA&5541&0.346&-0.369&3.219 &0.031&0.106&29.683&3\\
9&Mayer to USA&5546&0.393&-0.394&4.343 &0.026&-0.182&33.918&3\\
10&Olmeccu to USA&5551&0.342&-0.397&3.454 &0.027&-0.258&36.413&3\\
11&Ekofisk&5620&0.189&-0.168&3.374 &0.023&-0.018&7.875&5\\
12&Flotta&5602&0.209&-0.177&3.425 &0.024&-0.015&8.289&5\\
13&Forties&5630&0.204&-0.394&2.588 &0.023&-0.872&22.102&5\\
14&Gullfaks&5148&0.134&-0.167&3.278 &0.023&-0.224&7.432&5\\
15&Oseberg&5929&0.185&-0.168&3.693 &0.023&-0.051&8.480&5\\
16&Statfjord&5601&0.187&-0.167&3.347 &0.023&-0.055&8.132&5\\
17&Urals-Meaditerrean&5442&0.168&-0.175&3.266 &0.024&-0.230&7.724&5\\
18&Urals-North West Europe&5450&0.256&-0.178&3.242 &0.024&0.074&9.564&5\\
19&Dated BFOE&7514&0.278&-0.166&2.322 &0.023&0.069&10.865&5\\
20&Saharan Blend&5602&0.201&-0.390&2.486 &0.023&-0.919&22.192&5\\
21&Zaraitine&5442&0.131&-0.166&3.134 &0.022&-0.170&6.973&5\\
22&Swez Blend&5603&0.376&-0.260&3.751 &0.027&0.259&15.369&5\\
23&Amna&5579&0.247&-0.267&3.302 &0.024&-0.145&12.050&5\\
24&Brega&5603&0.258&-0.399&3.118 &0.023&-0.698&25.250&5\\
25&Es Sider&5603&0.241&-0.171&3.393 &0.023&0.043&9.169&5\\
26&Sarir&5603&0.249&-0.170&3.461 &0.023&0.029&9.651&5\\
27&Sirtica&5579&0.237&-0.256&3.122 &0.023&-0.143&11.746&5\\
28&Zueitina&5581&0.241&-0.260&2.887 &0.023&-0.183&12.216&5\\
29&Bonny Light&6709&0.177&-0.398&2.836 &0.023&-0.922&22.354&5\\
30&Brass River&5604&0.165&-0.398&2.604 &0.023&-1.026&23.562&5\\
31&Escravos&5583&0.241&-0.260&3.222 &0.023&-0.208&12.184&5\\
32&Forcados&6196&0.170&-0.166&3.952 &0.023&-0.160&8.337&5\\
33&Pennington&5587&0.246&-0.265&3.338 &0.022&-0.216&12.943&5\\
34&Kole&5603&0.220&-0.185&3.482 &0.024&-0.105&9.113&5\\
35&Lokele&5603&0.294&-0.195&3.593 &0.025&0.073&11.210&5\\
36&Mandji&5602&0.279&-0.197&3.801 &0.025&0.052&10.349&5\\
37&Djeno&5603&0.232&-0.185&3.750 &0.025&-0.040&8.639&5\\
38&Caninda&5583&0.248&-0.177&3.647 &0.024&-0.020&9.214&5\\
39&Gippsland&5562&0.205&-0.250&2.575 &0.020&-0.527&14.813&1\\
40&Daqing&5562&0.165&-0.235&2.727 &0.021&-0.457&14.207&1\\
41&Labuan OSP&5512&0.151&-0.148&3.254 &0.019&-0.193&7.583&5\\
42&Tapis&5560&0.152&-0.251&2.627 &0.019&-0.575&13.096&1\\
43&Attaka&5562&0.336&-0.333&2.679 &0.021&-0.579&36.087&1\\
44&Cinta&5561&0.200&-0.235&2.766 &0.022&-0.526&14.640&1\\
45&Duri&5562&0.388&-0.383&2.883 &0.024&-0.290&32.310&1\\
46&Lalang&5556&0.147&-0.215&2.712 &0.020&-0.490&11.216&1\\
47&Minas&5562&0.179&-0.264&2.719 &0.021&-0.694&14.826&1\\
48&Arab Light to Asia&16167&0.396&-0.232&2.554 &0.017&1.519&52.236&4\\
49&Arab Light to Europe&5651&0.373&-0.292&4.248 &0.025&0.500&30.668&6\\
50&Arab Light to USA&5572&0.372&-0.310&4.386 &0.028&0.285&19.165&3\\
51&Arab Medium to Asia&5646&0.358&-0.279&4.436 &0.026&0.220&17.152&5\\
52&Arab Medium to Europe&5643&0.396&-0.242&4.498 &0.025&0.711&24.659&6\\
53&Arab Medium to USA&5572&0.392&-0.314&4.652 &0.030&0.281&17.506&3\\
54&Arab Heavy to Asia&6129&0.296&-0.325&3.429 &0.029&-0.273&15.293&4\\
55&Arab Heavy to Europe&5644&0.253&-0.242&4.081 &0.026&0.062&15.672&6\\
56&Arab Heavy to USA&5572&0.372&-0.312&5.079 &0.032&0.229&16.258&3\\
57&Dubai Fateh&7669&0.276&-0.188&2.179 &0.023&0.277&13.695&2\\
58&Murban&5579&0.233&-0.168&3.675 &0.022&0.153&9.958&2\\
59&Lower Zakum&5579&0.233&-0.170&3.680 &0.023&0.051&10.022&2\\
60&Upper Zakum&5576&0.255&-0.173&3.749 &0.024&0.154&10.680&2\\
61&Oman&6113&0.246&-0.162&4.179 &0.024&0.172&10.103&2\\
62&Qatar Land&5578&0.291&-0.164&3.715 &0.023&0.360&13.222&2\\
63&Qatar Marine&5577&0.302&-0.172&3.635 &0.024&0.393&13.783&2\\
64&Syria Light&5647&0.376&-0.315&4.094 &0.025&0.173&23.050&5\\
65&Kuwait to Asia&5647&0.379&-0.259&3.592 &0.027&0.421&18.108&4\\
66&Iran Light to Asia&5644&0.324&-0.214&3.531 &0.025&0.274&15.397&4\\
67&Iran Light to NWE&5644&0.362&-0.300&4.221 &0.024&0.566&28.401&6\\
68&Iran Light to SKirr&5618&0.163&-0.197&3.522 &0.023&-0.040&12.201&6\\
69&Iran Heavy to Asia&5644&0.268&-0.291&2.836 &0.026&-0.196&13.014&4\\
70&Iran Heavy to NWE&5643&0.280&-0.264&3.821 &0.024&-0.064&18.284&6\\
71&Iran Heavy to Skirr&5617&0.170&-0.199&3.526 &0.024&0.010&11.508&6\\
\hline
\hline
\end{longtable}
\end{center}

\end{document}